\documentclass[preprint,NumberedRefs]{JASA}
\usepackage{amsmath}
\usepackage{amssymb}
\usepackage{graphicx}
\usepackage{color}
\usepackage{bm}
\usepackage{epsfig,psfrag}
\usepackage{tabularx}
\usepackage{acronym}
\usepackage[yyyymmdd,hhmmss]{datetime}
\usepackage{bbm}
\usepackage{mathrsfs} 
\usepackage{bardiag}
\usepackage{tabularx}
\usepackage{wrapfig}
\usepackage{multirow}
\usepackage{colortbl,pgfplotstable}
\usepackage{LatexInclusion/notation}
\usepackage[level=3]{LatexInclusion/messages}  
\usepackage{caption}
\usepackage{subcaption}
\usepackage{overpic}

\newcommand{\bd}{\begin{description}}
\newcommand{\ed}{\end{description}}
\newcommand{\be}{\begin{enumerate}}
\newcommand{\ee}{\end{enumerate}}
\newcommand{\bi}{\begin{itemize}}
\newcommand{\ei}{\end{itemize}}
\newcommand{\bl}{\begin{list}}
\newcommand{\el}{\end{list}}
\newcommand{\bt}{\begin{tabbing}}
\newcommand{\et}{\end{tabbing}}

\definecolor{BLUE}{rgb}{0,0,1}

\allowdisplaybreaks
\definecolor{BLUE}{rgb}{0,0,1}
\definecolor{BLACK}{rgb}{0,0,0}

\renewcommand{\bl}[1]{\textcolor{blue}{#1}}

\providecommand{\ist}{\hspace*{.3mm}}

\providecommand{\rmv}{\hspace*{-.3mm}}

\providecommand{\nn}{\nonumber}

\acrodef{pnt}[PNT]{positioning, navigation and timing}
\acrodef{iot}[IoT]{Internet of Things}
\acrodef{d2d}[D2D]{device-to-device}
\acrodef{aoa}[AOA]{angle-of-arrival}
\acrodef{nls}[NLS]{network localization and synchronization}
\acrodef{bp}[BP]{belief propagation}
\acrodef{spbp}[SPBP]{sigma point belief propagation}
\acrodef{fg}[FG]{factor graph}
\acrodef{pdf}[PDF]{probability density function}
\acrodef{mmse}[MMSE]{minimum mean-square error}
\acrodef{sp}[SP]{sigma point}
\acrodef{trwspa}[TRW-SPA]{{tree-reweighted sum-product algorithm}}
\acrodef{armse}[ARMSE]{average root mean square error}
\acrodef{uwb}[UWB]{ultra-wide bandwidth}
\acrodef{gnss}[GNSS]{global navigation satellite system}
\acrodef{spa}[SPA]{sum-product algorithm}
\acrodef{trw}[TRW]{{tree-reweighted}}
\acrodef{eap}[EAP]{edge appearance probability}
\acrodef{pmf}[PMF]{probability mass function}
\acrodef{eop}[EOP]{error outage probability}
\acrodef{ospa}[OSPA]{optimal subpattern assignment}
\acrodef{gospa}[GOSPA]{generalized optimal sub-pattern assignment}
\acrodefplural{eap}[EAPs]{edge appearance probabilities}

\acrodef{pt}[PT]{potential target}
\acrodef{spada}[SPADA]{sum-product algorithm for data association}
\acrodef{da}[DA]{data association}
\acrodef{mot}[MOT]{multiobject tracking}
\acrodef{mtt}[MTT]{multi-target tracking}
\acrodef{eot}[EOT]{extended object tracking}
\acrodef{pda}[PDA]{probabilistic data association}
\acrodef{jpda}[JPDA]{Joint \ac{pda}}
\acrodef{phd}[PHD]{probability hypothesis density}
\acrodef{cphd}[CPHD]{cardinalized \ac{phd}}
\acrodef{gmphd}[GM-PHD]{Gaussian mixture probability hypothesis density}
\acrodef{mht}[MHT]{multi-hypothesis tracking}
\acrodef{rfs}[RFS]{random finite set}
\acrodef{slam}[SLAM]{simultaneous localization and mapping}
\acrodef{iid}[iid]{independent and identically distributed}
 
\acrodef{pam}[PAM]{passive acoustic monitoring}
\acrodef{doa}[DOA]{direction of arrival}
\acrodef{tdoa}[TDOA]{time difference of arrival}
\acrodef{tsa}[TSA]{track segment association}
\acrodef{harp}[HARP]{High-frequency Acoustic Recording Package}
\acrodef{snr}[SNR]{signal-to-noise ratio}
\acrodef{ici}[ICI]{inter-click interval}
\acrodef{drtd}[DRTD]{Direct Reflected Time Differences}

\definecolor{BLUE}{rgb}{0,0,1}

\definecolor{myred}{rgb}{1,0.27,0}
\definecolor{mygreen}{rgb}{0.1, 0.55, 0.1}
\definecolor{myblue}{rgb}{0, 0, 1}

\definecolor{red}{rgb}{1,0,0}
\colorlet{RED}{red}

\newdateformat{monthyeardate}{\monthname[\THEMONTH] \THEDAY, \THEYEAR} 

\newcommand{\paperTitleMarkboth}{}

\pgfplotsset{compat=1.14}

\makeatletter
\expandafter\let\csname longtable*\endcsname\relax
\expandafter\let\csname endlongtable*\endcsname\relax
\makeatother

\begin{document}

\title{Bridging Echolocation Gaps in Automated Beaked Whale Tracking\vspace{0mm}}

\author{Clair Ma}
\affiliation{Scripps Institution of Oceanography and Department of Electrical and Computer Engineering, University of California at San Diego, La Jolla, California 92093, USA}
\email{c2ma@ucsd.edu}
\author{Thomas Kropfreiter}
\affiliation{Institute of Telecommunications, TU Wien, 1040 Vienna, Austria}
\author{Lauren Baggett}
\affiliation{Scripps Institution of Oceanography, University of California at San Diego, La Jolla, California 92093, USA.}
\author{Simone Baumann-Pickering}
\affiliation{Scripps Institution of Oceanography, University of California at San Diego, La Jolla, California 92093, USA.}
\author{Florian Meyer}
\affiliation{Scripps Institution of Oceanography and Department of Electrical and Computer Engineering, University of California at San Diego, La Jolla, California 92093, USA}

\begin{abstract}
{Passive acoustic monitoring (PAM) is an effective and widely used tool for tracking marine mammals, particularly beaked whales, which are infrequently observed visually because of their deep-diving behavior.} 
However, the large data sets generated by PAM methods often require time-consuming hand labeling to identify whale trajectories in the recorded audio. 
{Automated multi-target tracking (MTT) methods could significantly reduce human workload, but current methods have difficulty forming continuous tracks because of the irregularity of beaked whale echolocation clicks.} 
{More precisely, regular sequences of clicks are often interrupted by longer pauses that occur when whales face away from the sensors or stop clicking.} 
{Consequently, the probability of detection is difficult to model accurately, and MTT trajectories become fragmented at these pauses.}
{In this paper, we propose a multistage target-estimation method aimed at bridging large gaps in click sequences by combining belief propagation-based MTT with track smoothing and stitching.} 
{We validate our method using acoustic recordings of clicks from goose-beaked whales (\textit{Ziphius cavirostris}), and demonstrate that it improves track estimates and reduces fragmentation in the presence of consecutive missed detections. When evaluated with the generalized optimal subpattern assignment (GOSPA) metric, our method outperforms existing MTT reference methods through reductions in missed-target errors.}

\end{abstract}

\maketitle


\acresetall
\section{Introduction}
\label{sec:introduction}
\Ac{pam} is a method that is well-suited for marine ecological monitoring due to the long-distance propagation of acoustic waves in underwater environments and the ability of modern acoustic sensors to be deployed over longer periods of time.
In particular, \ac{pam} is useful for studying marine species that frequently produce sounds, such as whales (cetaceans), for which methods like visual surveys and tagging may be difficult or ineffective \cite{Zim:B11,MelStaMooDziMat:J07}.  
Whales produce a range of sounds depending on species and function, such as the frequency modulated upswept echolocation pulses used by beaked whales for foraging \cite{BauMcDSimSolMerOleRocWigRanYacHil:J13,JohMadZimAguTya:J04}.
Through \ac{pam} methods, these signals can provide insights into whale {behavior} and population numbers in a particular region.

Beaked whales, or toothed whales belonging to the family \textit{Ziphiidae}, are of particular scientific interest due to their sensitivity to intense anthropogenic sound, having been observed in several lethal mass strandings associated with naval mid-frequency active sonar events \cite{CoxRagReaVosBaiBalBarCalCraCru:J06,SimLop:R91,SimBroThaTriOleHunBau:R20,FilMinMicDamTyaKet:R09}. Because of their deep-diving behavior and rarity in visual surveys \cite{TyaJohAguStuMad:R06,SchFalMorAnd:R14}, these whales are difficult to study through traditional means. While foraging in deep sea environments, beaked whales produce trains of echolocation clicks that help them locate prey and navigate in darkness \cite{JohMadZimAguTya:J04}. Sound therefore plays a significant role in the behavior of beaked whales, which can be investigated through \ac{pam} methods. 

After recording beaked whale clicks, localization can be performed for each click, providing insight into the whales' diving and foraging behavior, as well as the number of individuals \cite{BarGriKliHar:J18,GasWigHil:J15,BagSnyBerCurWigHilSirFraBau:J25,JanMeySnyWigBauHil:J23,SnySolSimFraWigHil:J24}. 
However, since beaked whales frequently dive in small groups \cite{BagSnyBerCurWigHilSirFraBau:J25,AlcJohArrWarPerMarMadAgu:J21}, there often exists ambiguity about which whale generated each click detection.
Beaked whales also produce highly directional echolocation clicks \cite{ZimJohMadTya:J05}, resulting in missed detections when whales face away from the sensors \cite{ShaMorJarTyaJoh:J13}. 
Furthermore, the potentially large volume of data produced by \ac{pam} methods makes data processing a nontrivial problem. As reviewed by Van Parijs et al.\cite{VanParijs2009}, archival and real-time \ac{pam} systems operate over broad temporal and spatial scales, and archival systems can generate data sets too large for manual inspection, making automated processing essential. In this work, we analyze recordings of \textit{Ziphius cavirostris}, commonly known as goose-beaked whales, collected using bottom-mounted volumetric hydrophone arrays, and demonstrate a method for automated labeling and 2-D localization of beaked whale echolocation clicks using a \ac{bp}-based \ac{mtt} algorithm.

\subsection{State of the Art}
Various types of \ac{pam} technology have been applied to whale tracking, including dark fiber-optic cables \cite{BouTawKriRorPotLanJohBreHauSchSto:J22}, directional acoustic vector sensors \cite{ThoSkiScoRosStrFol:J10,TenThoLamConKim:J22,GruJanKugKroTenLamThoMey:J23}, and arrays of omnidirectional hydrophones that are towed \cite{GruNosOle:J21}, mobile \cite{FreHarMatMelBarBauKli:J20}, or bottom-mounted \cite{GasWigHil:J15,BagSnyBerCurWigHilSirFraBau:J25}. For hydrophone arrays, an established method of click detection involves cross-correlating the acoustic signals from pairs of hydrophones to identify peaks. Each click can then be localized by extracting the \ac{tdoa} between hydrophones. Depending on the number of hydrophone pairs, these \ac{tdoa} measurements can be used to compute a whale's 2- or 3-D position \cite{Zim:B11}. 

Traditionally, trained human operators or machine learning methods are used to identify time windows containing clicks from the species of interest \cite{ZieFraHilOleBaiWigBau:J22}. 
A human operator then visually examines the click sequences to determine the number and trajectories of whales.
{Software toolkits can perform cross-correlation to obtain \acp{tdoa}, estimate the bearing of individual echolocation clicks, and provide a graphical user interface that displays the information for convenient labeling \cite{SnySolSimFraWigHil:J24}. However, a human operator is often still required to label each detection manually.}

Automated \ac{mtt} offers an alternative to the conventional hand-labeling approach and has the potential to be less time- and labor-intensive. 
The goal of \ac{mtt} is to estimate the time-dependent number and states of multiple targets in the presence of noisy and cluttered sensor measurements, missed detections, and measurement origin uncertainty \cite{BarWilTia:B11,Mah:B07,ChaMor:B11,MeyKroWilLauHlaBraWin:J18}. 
Various \ac{mtt} methods have been developed and can be categorized into ``vector-based'' and ``set-based'' methods. 
Vector-based methods model target states using random vectors. These include methods based on \ac{pda} \cite{ForBarSch:83,BarWilTia:B11,MusEva:J04}, {\ac{mht}} \cite{Rei:J79,Bla:J04}, and graphical models \cite{MeyKroWilLauHlaBraWin:J18,MeyWil:J21}. 
Set-based methods, on the other hand, model the target states using \ac{rfs}. These include methods based on unlabeled \acp{rfs} \cite{Mah:J03,Mah:J07,VoVoCan:J09}, such as the \ac{phd} filter \cite{Mah:J03}, and methods based on labeled \acp{rfs} \cite{VoVo:J13,ReuVoVoDie:J14,KroMeyHla:20,VoVoNguShi:J24}.
The \ac{bp}-based multisensor \ac{mtt} framework most directly related to this paper was developed in \onlinecite{MeyBraWilHla:15,MeyBraWilHla:J17,MeyKroWilLauHlaBraWin:J18}. \ac{bp}-based multisensor \ac{mtt} is particularly suited for nonlinear problems with multiple sensors, but also has moderate performance advantages compared to reference methods in linear problems with a single sensor \cite{MeyBraWilHla:15,MeyBraWilHla:J17,MeyKroWilLauHlaBraWin:J18}. In this work, the multisensor \ac{mtt} framework is combined with smoothing and stitching to address track fragmentation caused by extended detection gaps.

Track stitching is closely related to track-to-track association, which is a prerequisite to track-to-track fusion in distributed multisensor systems \cite{BarLi:95,BarWilTia:B11,TiaBar:J09,Gov:B23}. In the classical setting, contemporaneous tracks reported by different local trackers are associated by target identity after their state estimates have been expressed at a common time; the associated state estimates can then be fused. Because the local estimates may contain common information, the fusion step must account for their cross-correlation \cite{Bar:J81,GovKoc:J12}. The problem considered here is temporally complementary: we associate non-overlapping fragments produced by one tracker before and after a detection gap and concatenate their histories, rather than fusing concurrent state estimates.

Similar methods of target localization and tracking from acoustic arrays have been previously applied outside of the whale tracking problem. The early four-hydrophone system of Watkins and Schevill\cite{watkins71} demonstrated 3-D source localization from arrival-time differences with a non-rigid array. It provided accurate locations for nearby sources but primarily bearing information, with increasingly uncertain range, for distant sources.
In an earlier passive acoustic application, Meyer and Tesei\cite{MeyTes:R16} formulated the detection, localization, and tracking of an unknown number of sources from pairwise \ac{tdoa} measurements in a Bayesian framework. Their particle-based method accounts for the nonlinear measurement model, clutter, missed detections, and measurement-origin uncertainty; although it was evaluated on surface-vessel data rather than whale recordings, it addresses several of the same inference challenges as the present work.

In the following, we {review} \ac{mtt} methods for automated whale tracking.
{For wide-baseline arrays, Nosal\cite{Nos:J13} formed model-based spatial likelihoods from candidate \ac{tdoa} or time-of-arrival measurements and separated animals using persistent position-domain peaks. This avoids explicit association of \acp{tdoa} across hydrophone pairs and the prior removal of candidates caused by multipath or mismatched arrivals. The method was demonstrated on sperm whale recordings, but it does not explicitly reconnect track fragments across extended detection gaps.}
{The automated pipeline of Helble et al.\cite{Helble2015} combines spectral templating, cross-correlation of multi-unit sequences, model-based \ac{tdoa} localization, and track-based post-processing to localize multiple humpback whales and assign song units to individual singers on a wide-baseline range. The multi-unit sequences help reject ambiguous call matches and enable faster-than-real-time processing, but valid localizations rely on detections across the hydrophones of a subarray and the method does not address continuity through extended detection gaps.}
The approach used by Baggenstoss\cite{Bag:J15} tracks beaked whales by first performing target localization using \acp{tdoa} to obtain {3-D} position measurements, and then applying a \ac{mht} method to these measurements.
{The method by Gerard et al. \cite{GerCarCor:C08} also applies \ac{mht} to clicks produced by beaked whales; however, it estimates the number of whales present but not their positions over time and therefore performs tracking in the amplitude domain.} 
The \ac{gmphd} filter used by Gruden, Nosal, and Oleson \cite{GruNosOle:J21} to track false killer whales takes into account not only the \ac{tdoa} measurements of both broadband clicks and narrowband whistles, but also amplitude information. 
{The two-stage approach from Gruden, Nosal, and Henderson \cite{GruNosHen:J25} first uses the \ac{gmphd} filter to form tracks of sperm whale clicks in the \ac{tdoa} domain and then uses these intermediate estimates as inputs to a second \ac{gmphd} stage that produces tracks in the spatial domain.}
{The method for beaked whale tracking proposed by Jang et al. \cite{JanMeySnyWigBauHil:J23} uses two \ac{bp}-based \ac{mtt} stages: the first tracks in the \ac{tdoa} domain, and the second fuses the \ac{tdoa} estimates for tracking in 3-D.}
{This approach tracks directly in 3-D from \ac{tdoa} measurements without an intermediate localization step. The resulting measurement model is nonlinear and is implemented using particle flow.}
The fully automated vector sensor approach by Gruden, Jang, et al. \cite{GruJanKugKroTenLamThoMey:J23} compares a \ac{gmphd} filter with a \ac{bp}-based \ac{mtt} method for azimuthal tracking of multiple singing humpback whales, demonstrating moderate performance advantages of BP.
Gruden, Jang, et al. \cite{GruJanKugKroTenLamThoMey:J23} also report that longer gaps in humpback whale song can fragment a single trajectory into multiple tracks and identifies joining the resulting 2-D track fragments as future work. More generally, existing automated marine mammal tracking methods do not provide a dedicated mechanism for reconnecting fragments separated by the prolonged, behavior-driven detection gaps considered here. The explicit probabilistic stitching stage proposed in this paper addresses this continuity problem.

\subsection{Contributions, Paper Organization, and Notations}
{This paper aims to reduce the time- and labor-intensive burden of manually labeling \ac{pam} data for marine mammal research by proposing an automated method that can handle complex whale behavior. In particular, our method combines \ac{pda}-based \ac{mtt} with track smoothing and stitching to estimate continuous target tracks in the 2-D azimuth--elevation domain.} 
{Under standard \ac{mtt} assumptions, given that a target is present, its detection at each time step is assumed to be an independent and identically distributed random event that depends only on the \ac{snr}.} {However, the probability of detecting a beaked whale using \ac{pam} depends on a variety of additional factors, such as the signal beamwidth, the target range, and the whale's dive behavior, which affects both the click rate and the signal direction \cite{HilBauFraTriMerWigMcdGarHarMarTho:J15}.} {Because these behaviors are neither necessarily random nor independent over time, typical \ac{mtt} methods do not account for the resulting gaps in target detections; estimated tracks are therefore terminated prematurely and later reinitialized.} {A track stitching post-processing stage can mitigate this fragmentation. In addition, track smoothing provides more accurate track state estimates than \ac{mtt} alone and may thereby improve the stitching results.}

{This work introduces a three-step process for tracking whales from their echolocation-click \ac{doa} measurements. First, we perform \ac{mtt} using a \ac{bp}-based algorithm that outputs a list of estimated tracks. These tracks are likely to be fragments of the true whale trajectories, with breaks where target detections are missing because a whale has either stopped clicking or turned away from the receivers. Next, track smoothing reduces track estimation errors caused by measurement noise. Finally, track stitching connects track segments by predicting the endpoint of each earlier segment forward to the starting time of each new segment and finding the lowest-cost pairings. In this way, we can estimate a continuous track for each observed whale more accurately, even when gaps occur in the click measurements. We demonstrate our method using passive acoustic recordings of \textit{Ziphius cavirostris} and evaluate the results against reference labels obtained by trained human operators.} 

Conceptually, our stitching stage transfers the common-origin test from classical track-to-track association to a temporal setting: it associates an old and a new fragment after predicting their state estimates to a common time \cite{BarLi:95,TiaBar:J09,Gov:B23}. Unlike conventional track-to-track fusion, it concatenates disjoint track histories rather than combining concurrent state estimates.

The key contributions of this paper are summarized as follows.
\begin{itemize}
\item We establish a three-stage method for tracking whales that combines \ac{bp}-based \ac{mtt} with track smoothing and stitching.
\item We motivate our track smoothing and stitching equations using the probabilistic model underlying the \ac{bp}-based \ac{mtt}.
\item {We evaluate the accuracy of our method using passive acoustic recordings of \textit{Ziphius cavirostris} echolocation clicks in five examples that represent varying levels of complexity in real data.}
\end{itemize}

This paper advances over the preliminary account of our method provided in the conference publication \cite{MaKroBagBauMey:C25} by (i) incorporating a factor graph representation of track smoothing, (ii) establishing the effectiveness of the method on simulated trials, (iii) testing on a wide range of beaked whale data reflective of the variation in real-world encounters, (iv) quantifying error through \ac{gospa} metrics, (vi) demonstrating performance advantages compared to state-of-the-art reference methods.

\emph{Notation:} Random variables are displayed in sans serif, upright fonts; their realizations in serif, italic fonts. 
Vectors and matrices are denoted by bold lowercase and uppercase letters, respectively. For example, a random variable and its realization are denoted by $\rv x$ and $x$, respectively, and a random vector and its realization 
by $\RV x$ and $\V x$, respectively. 
Furthermore, {${\V{x}}^{\mathrm{T}}$} denotes the transpose of vector $\V x$; 
$\propto$ indicates equality up to a normalization factor;
$f(\V x)$ denotes the \ac{pdf} of 
random vector $\RV x$ (this is a short notation for  $f_{\RV x}(\V x)$); 
$f(\V x | \V y)$ denotes the conditional \ac{pdf}  
of 
random vector $\RV x$ conditioned on random vector  $\RV y$  (this is a short notation for  {$f_{\RV x | \RV y}(\V x | \V y)$}); 
$\M I_n$ denotes the {$n \rmv\times\rmv n$ identity matrix}.

\section{BP-Based MTT}\label{sec:systemModel}
The central objective of performing \ac{mtt} is to estimate the number and states of multiple targets $\RV{x}_k^{(j)}$\rmv, $j\in\{1,\ldots,j_k\}$, at each time step $k \in \{1,\dots,K\}$. Here, $j_k$ {is} the number of targets at time step $k$. 
For the whale tracking problem, $\RV{x}_k^{(j)}$ will be a {4-D} vector representing azimuth, elevation, azimuthal velocity, and elevation velocity of the target/whale relative to the deployed sensor. 
We represent all target states at time step $k$ by the joint vector $\RV{x}_k \triangleq [\RV{x}_k^{(1)\mathrm{T}}\rmv\ldots\ist\RV{x}_k^{(j_k)\mathrm{T}}]^{\mathrm{T}}$.
 
At each step $k$, a sensor produces $m_k$ measurements $\V{z}_k^{(m)}$\rmv, $m \in \{1,\ldots,m_k\}$,
{where $\V{z}_k \triangleq [\V{z}_k^{(1)\mathrm{T}}\rmv\rmv\ldots\ist\ist\V{z}_k^{(m_k)\mathrm{T}}]^{\mathrm{T}}$ is the vector containing all sensor measurements collected at time $k$. In our whale tracking scenario, $\V{z}_k^{(m)}$ will be a 2-D vector representing noisy observations of the azimuth and elevation angles.} 
We further define $\V{z}_{1:k} \triangleq [\V{z}_1^{\mathrm{T}}\rmv\ldots\ist\V{z}_k^{\mathrm{T}}]^{\mathrm{T}}$ to represent all sensor measurements collected up to time $k$. 
{We consider a single-sensor scenario.}

{Our \ac{mtt} approach is based on the concept of \acp{pt}.}
Each \ac{pt} is represented by the vector $\RV{y}_k^{(j)}\rmv=[\RV{x}_k^{(j)\mathrm{T}} \rv{r}_k^{(j)}]^{\mathrm{T}}$\rmv, where 
$\rv{r}_k^{({j})} \rmv\in\rmv \{0,1\}$ is a binary existence variable that describes whether the corresponding \ac{pt} actually exists $(\rv{r}_k^{({j})} = 1)$ or not $(\rv{r}_k^{({j})} = 0)$.
{To estimate $\RV{x}_k$, we compute the marginal posterior \acp{pdf} $f(\V{x}_k^{(j)}\rmv,r_k^{(j)}|\V{z}_{1:k})$, $j\rmv\in\rmv\{1,\ldots,j_k\}$.} 
Based on $f(\V{x}_k^{(j)}\rmv,r_k^{({j})}|\V{z}_{1:k})$, we can further compute the marginal posterior \ac{pmf} $p(r_k^{({j})}|\V{z}_{1:k})$ according to 
\begin{equation}\label{eq:pePMF}
p(r_k^{(j)}|\V{z}_{1:k}) = \int f(\V{x}_k^{(j)}\rmv,r_k^{(j)}|\V{z}_{1:k})\ist\mathrm{d}\V{x}_k^{(j)}\ist.
\end{equation}
Here, $p(r_k^{(j)}|\V{z}_{1:k})$ represents the posterior probability that {the} \ac{pt} with state $\RV{y}_k^{(j)}$ really exists, i.e., represents an actual target.
{The quantity $p(r_k^{(j)}|\V{z}_{1:k})$ is therefore often referred to as the existence probability.} 
A \ac{pt} is {considered} to exist if $p(r_k^{(j)}|\V{z}_{1:k})$ is greater than a predefined threshold $\gamma_{\text{D}}$.  
For each existing \ac{pt}, we perform \ac{mmse} state estimation according to 
\begin{equation}
\label{eq:mmse}
\hat{\V{x}}_k^{(j)} \triangleq \int \V{x}_k^{(j)}{f(\V{x}_k^{(j)}|r_k^{(j)}=1\ist,\V{z}_{1:k})}\ist\mathrm{d}\V{x}_k^{(j)}
\end{equation}
where the conditional \ac{pdf} $f(\V{x}_k^{(j)}|r_k^{(j)}=1,\V{z}_{1:k})$ can in turn {be} calculated according to
\begin{equation}\label{eq:condPost}
f(\V{x}_k^{(j)}|r_k^{(j)}=1,\V{z}_{1:k}) = \frac{f(\V{x}_k^{(j)}\rmv,r_k^{(j)} \rmv=\rmv 1|\V{z}_{1:k})}{p(r_k^{(j)} \rmv=\rmv 1|\V{z}_{1:k})}\ist.
\end{equation}
{To compute \eqref{eq:pePMF}--\eqref{eq:condPost}, we use a \ac{bp}-based \ac{mtt} approach \cite{MeyBraWilHla:J17,MeyKroWilLauHlaBraWin:J18,KroMeyCroCorHlaWil:J24}.} 
The underlying statistical model will be described in the following subsection. 
{State estimation at time $k$ using past and current measurements $\V{z}_{1:k}$ constitutes a filtering problem, as encountered in real-time tracking. During post-processing, whale tracking can instead be formulated as a smoothing problem, in which the state at time $k$ is estimated using past, current, and future measurements $\V{z}_{1:K}$, where $K$ is the final time step in the tracking scenario.}

\subsection{Statistical Model}\label{sec:statMod}

Our \ac{bp}-based tracking method is based on the following model assumptions. At time $k-1$, a target with state $\RV{x}_{k-1}$ survives with probability $p_{\text{s}}$ to time $k$, at which its state $\RV{x}_{k}$ is distributed according to the state-transition \ac{pdf} $f(\V{x}_k|\V{x}_{k-1})$.
Otherwise, it dies with probability $1-p_{\text{s}}$.
{Newborn targets are modeled by a Poisson point process with mean parameter $\mu_{\text{b}}$ and spatial \ac{pdf} $f_{\text{b}}(\V{x}_{k})$. In our algorithm, we account for target birth and death using ``legacy'' and ``new'' \acp{pt}.} 
Surviving targets are represented by legacy \acp{pt} ${\RV{\underline{y}}}_k^{(j)}$, $j\in\{1,\ldots,j_{k-1}\}$ and newborn targets by new \acp{pt} $\RV{\overline{y}}_k^{(m)}$, $m\in\{1,\ldots,m_k\}$. 
For each measurement $m$, $m\in\{1,\ldots,m_k\}$, a single new \ac{pt} is generated to account for the possibility that the corresponding measurement was produced by a new, previously unobserved target. Furthermore, a new \ac{pt} generated in time step $k$ becomes a legacy \ac{pt} in the subsequent time step $k+1$ \cite{MeyKroWilLauHlaBraWin:J18}.
The vector $\RV{y}_k$ collects all legacy and new \ac{pt} states at time $k$, i.e., $\RV{y}_k=[\underline{\RV{y}}_k^{(1)\mathrm{T}}\rmv\rmv\ldots\ist\ist\underline{\RV{y}}_k^{(j_{k-1})\mathrm{T}}\ist \overline{\RV{y}}_k^{(j_{k-1}+1)\mathrm{T}}\ldots\overline{\RV{y}}_k^{(j_{k-1}+m_k)\mathrm{T}}]^\mathrm{T}$ with $\underline{\RV{y}}_k^{(j)} = [\underline{\RV{x}}_k^{(j)\mathrm{T}}\ist \underline{\rv{r}}_k^{(j)}]^{\mathrm{T}}$ and $\overline{\RV{y}}_k^{(m)} = [\overline{\RV{x}}_k^{(m)\mathrm{T}}\ist \overline{\rv{r}}_k^{(m)}]^{\mathrm{T}}$\rmv. 

{Each target $\RV{x}_k^{(j)}$ is detected by the sensor with probability $p_{\text{d}}$ and is not detected with probability $1 - p_{\text{d}}$.} 
If detected, a target generates exactly one measurement $\RV{z}_k^{(m)}$\rmv, which is distributed according to the conditional \ac{pdf} $f(\V{z}^{(m)}_{k} | \V{x}^{(j)}_{k})$.
There are also clutter measurements generated by non-target sources. 
Clutter measurements are modeled by a Poisson point process with mean parameter $\mu_{\text{c}}$ and spatial \ac{pdf} $f_{\text{c}}(\V{z}^{(m)}_{k})$. 
Since the origin of each measurement is unknown, i.e., we have no prior information on whether a measurement was generated by a given target or by clutter, we introduce the random ``target-oriented'' \ac{da} vector $\RV{a}_{k}$.
Here, $\RV{a}_{k} = \big[\rv{a}_{k}^{(1)} \rmv\cdots\ist \rv{a}_{k}^{(j_{k-1})} \big]^{\mathrm{T}}\rmv$, where $\rv{a}_{k}^{(j)} = m \in \{1,\dots,m_{k}\}$ means that legacy \ac{pt} $j$ generated measurement $m$, and $\rv{a}_{k}^{(j)} = 0$ means that legacy \ac{pt} $j$ did not produce a measurement.
For a valid one-to-one association, the nonzero entries of $\RV{a}_k$ are distinct. Measurement $m$ is assigned to the unique legacy \ac{pt} $j$ satisfying $\rv{a}_k^{(j)}=m$; if no such $j$ exists, it was generated by a new \ac{pt} or by clutter. Thus, whether a measurement is assigned to a legacy \ac{pt} is completely determined by $\RV{a}_k$ and need not be represented by a separate random variable.

Using these model assumptions, the corresponding joint posterior \ac{pdf} of $\RV{y}_{0:K}$ and $\RV{a}_{1:K}$ conditioned on $\V{z}_{1:K}$ can be written in the target-oriented form \cite{MeyKroWilLauHlaBraWin:J18}
\vspace{0mm}
\begin{align}
f(\V{y}_{0:K},\V{a}_{1:K}|\V{z}_{1:K}) &\propto \prod_{k=\ist 1}^{K} \Bigg\{ \Psi(\V{a}_{k})
\prod_{j'=\ist 1}^{j_{k-1}} f\big(\V{\underline{y}}_{k}^{(j')}|\V{y}_{k-1}^{(j')}\big) \nn\\[-0.5mm]
&\quad\times \prod_{j=\ist 1}^{j_{k-1}} q\big(\V{\underline{x}}_{k}^{(j)}\rmv,\underline{r}_{k}^{(j)}\rmv,a_{k}^{(j)};\V{z}_{k}\big)
\prod_{m=\ist 1}^{m_{k}} v\big(\V{\overline{x}}_{k}^{(m)}\rmv,\overline{r}_{k}^{(m)}\rmv,\V{a}_{k};\V{z}_{k}^{(m)}\big) \Bigg\}. \label{eq:jointPosteriorPDF}
\end{align}
Here, $f\big(\V{\underline{y}}_{k}^{(j)}|\V{y}_{k-1}^{(j)}\big)$ is referred to as {the} single-\ac{pt} state-transition \ac{pdf} and {is} given by
\vspace{-1mm}
\begin{equation}
 f\big( \underline{\V{x}}^{(j)}_{k}\rmv, \underline{r}^{(j)}_{k} \big| \V{x}^{(j)}_{k-1}, r^{(j)}_{k-1} = 0 \big) = 
   \begin{cases}
   f_{\text{D}}\big( \underline{\V{x}}^{(j)}_{k}\big) \ist, & \!\! \underline{r}^{(j)}_{k} \!\rmv=\! 0, \\[.8mm]
   0 \ist, & \!\! \underline{r}^{(j)}_{k} \!\rmv=\! 1, 
   \end{cases} 
	\vspace{-1mm}
\label{eq:singleTargetStateTrans_0}
\end{equation}
and 
\begin{align}
f\big( \underline{\V{x}}_{k}^{(j)}\rmv, \underline{r}_{k}^{(j)} \big| \V{x}_{k-1}^{(j)}, r^{(j)}_{k-1} = 1 \big) &=  
   \begin{cases}
		\big( 1 \!\rmv-\rmv p_{\text{s}} \big) \ist f_{\text{D}}\big( \underline{\V{x}}^{(j)}_{k}\big) \ist, & \!\!\underline{r}^{(j)}_{k} \!\rmv=\! 0, \\[1.5mm] 
		p_{\text{s}} \ist f\big( \underline{\V{x}}^{(j)}_{k} \big| \V{x}^{(j)}_{k-1}\big) \ist, & \!\!\underline{r}^{(j)}_{k} \!\rmv=\! 1. 
   \end{cases} \label{eq:singleTargetStateTrans_1}\\[-13mm]
	\nn
\end{align}
where $f_{\text{D}}\big( \V{x}^{(j)}_{k}\big)$ is an arbitrary dummy \ac{pdf}. 
{The single-target state-transition \ac{pdf} $f\big( \underline{\V{x}}^{(j)}_{k} \big| \V{x}^{(j)}_{k-1}\big)$ is modeled using a constant-velocity motion model \cite{BarWilTia:B11}, i.e.,}
\vspace{2mm}
\begin{equation}
\label{eq:stateTransitionModel}
\V{\underline{x}}_k^{(j)} = \begin{bmatrix}
1 & 0 & t_k & 0\\
0 & 1 & 0 & t_k\\
0 & 0 & 1 & 0\\
0 & 0 & 0 & 1
\end{bmatrix}
\V{\underline{x}}_{k-1}^{(j)} + \begin{bmatrix}
0.5t_k^2 & 0\\
0 & 0.5t_k^2\\
t_k & 0\\
0 & t_k\\
\end{bmatrix}\V{w}_k^{(j)}
\end{equation}
{where $\V{w}_k^{(j)} \rmv\sim\rmv \mathcal{N}(\V{w}_k^{(j)};\V{0},\sigma_\mathrm{w}^2\M{I}_2)$ is the driving noise, $\sigma_\mathrm{w}^2$ is the driving noise variance, and $t_k$ is the interval between two consecutive time steps.} 
{If an irregular time grid is used, the survival probability $p_{\text{s}}$ in \eqref{eq:singleTargetStateTrans_1} may depend on $t_k$. Furthermore, a larger value of $\sigma_\mathrm{w}^2$ can account for possible model mismatch resulting from applying the constant-velocity motion model in the angular domain.} 

The factor $q\big( \underline{\V{x}}^{(j)}_{k}\!, \underline{r}^{(j)}_{k}\!, a^{(j)}_{k} \rmv; \V{z}_{k} \big)$ in \eqref{eq:jointPosteriorPDF} is determined by the measurement model. For $\underline{r}^{(j)}_{k} =0$, it is given by $q\big( \underline{\V{x}}^{(j)}_{k},\underline{r}^{(j)}_{k} =0, a^{(j)}_{k}\rmv; \V{z}_{k} \big) = 1(a^{(j)}_{k})$ with {$1(a)$} being an indicator function that is one for $a=0$ and zero otherwise, and, for $\underline{r}^{(j)}_{k} =1$, by
\begin{equation}
\label{eq:qFunction}
q\big( \underline{\V{x}}^{(j)}_{k}\!, \underline{r}^{(j)}_{k} =1, a^{(j)}_{k}\rmv; \V{z}_{k} \big) = \begin{cases}
    \frac{ p_{\text{d}} }{ \mu_{\text{c}}\ist f_{\text{c}}( \V{z}_{k}^{(m)} )} \ist f\big( \V{z}_{k}^{(m)} \rmv\big|\ist \underline{\V{x}}_{k}^{(j)} \big), & a^{(j)}_{k} \!=\rmv m \rmv\in\rmv \{1,\dots,m_k \} \\[2.5mm]
     1 \!-\! p_{\text{d}}  \ist, & a^{(j)}_{k} \!=\rmv 0\ist. 
  \end{cases} 
\end{equation}
The conditional {\ac{pdf}} $f\big( \V{z}_{k}^{(m)} \rmv\big|\ist \V{x}_{k}^{(j)} \big)$ is given by a linear/Gaussian model according to
\begin{equation}
\label{eq:MeasMod}
\V{z}_k^{(m)} = \begin{bmatrix}
1 & 0 & 0 & 0\\
0 & 1 & 0 & 0
\end{bmatrix}
\V{x}_{k}^{(j)} + \V{\epsilon}_k^{(m)}
\end{equation}
where $\V{\epsilon}_k^{(m)} \rmv\sim\rmv \mathcal{N}\big(\V{\epsilon}_k^{(m)};\V{0}_2,\M{I}_2 \ist[\sigma^2_{v,\text{a}} \sigma^2_{v,\text{e}}]^{\mathrm{T}}\big)$ is the measurement noise with azimuth variance $\sigma^2_{v,\text{a}}$ and elevation variance $\sigma^2_{v,\text{e}}$.

Next, the factor $v\big(\overline{\V{x}}^{(m)}_{k}\!,\overline{r}^{(m)}_{k}\!,\V{a}_k;\V{z}_{k}^{(m)}\big)$ represents the likelihood of a new target being the origin of an observed measurement. The factor depends on $\V{a}_k$ only to determine whether measurement $m$ has already been assigned to any legacy \ac{pt}. For $\overline{r}^{(m)}_{k}\rmv=0$, it is given by $v\big(\overline{\V{x}}^{(m)}_{k}\!,\overline{r}^{(m)}_{k}\rmv=0,\V{a}_k;\V{z}_{k}^{(m)}\big)\rmv=\rmv f_{\text{D}}\big(\overline{\V{x}}^{(m)}_{k}\big)$ and, for $\overline{r}^{(m)}_{k}\rmv=1$, by
\begin{equation}
\label{eq:vFunction}
v\big(\overline{\V{x}}^{(m)}_{k}\!,\overline{r}^{(m)}_{k}\rmv=1,\V{a}_{k}\rmv;\V{z}_{k}^{(m)}\big) =
\begin{cases}
0 \ist, & \exists j\in\{1,\ldots,j_{k-1}\}: a^{(j)}_{k}\!\rmv=\rmv m, \\[1mm]
\displaystyle \frac{p_{\text{d}} \ist \mu_{\text{b}} \ist f_{\text{b}}(\overline{\V{x}}^{(m)}_{k})}{\mu_{\text{c}} \ist f_{\text{c}}(\V{z}_{k}^{(m)})} \ist f\big(\V{z}_k^{(m)}\big|\overline{\V{x}}^{(m)}_{k}\big) \ist,
& \forall j\in\{1,\ldots,j_{k-1}\}: a^{(j)}_{k}\!\rmv\neq\rmv m.
\end{cases}
\end{equation}
Finally, $\Psi(\V{a}_k)$ is an indicator function that enforces a one-to-one association by prohibiting two legacy \acp{pt} from generating the same measurement and is defined as\vspace{-2mm}
\begin{equation}\label{eq:psi}
\Psi(\V{a}_k) \triangleq
\begin{cases}
0, & \exists j\neq j': a_k^{(j)}=a_k^{(j')}\neq 0,\quad j,j'\in\{1,\ldots,j_{k-1}\},\\[1mm]
1, & \text{otherwise}.
\end{cases}
\end{equation}
Having introduced the underlying statistical model and the corresponding joint posterior \ac{pdf} $f(\V{y}_{0:K},\V{a}_{1:K}|\V{z}_{1:K})$, we outline the \ac{bp}-based \ac{mtt} approach in the next subsection.

\subsection{Message Passing and Selected Processing Steps}\label{sec:MP}
The factorization of the joint posterior \ac{pdf} $f(\V{y}_{0:k},\V{a}_{1:k}|\V{z}_{1:k})$ in \eqref{eq:jointPosteriorPDF} admits a factor graph representation \cite{KolFri:B09,KscFreLoe:01}. For efficient inference, we use the equivalent scalable \ac{bp} message computations developed in \onlinecite{MeyKroWilLauHlaBraWin:J18,KroMeyCroCorHlaWil:J24}, where the complete factor graph construction is provided.
\ac{bp}, often also referred to as the sum-product algorithm, is a method that operates on a factor graph to efficiently compute marginal \acp{pdf}/\acp{pmf} \cite{KolFri:B09}. In the \ac{bp} algorithm, local operations are performed at individual graph nodes, and the results of these operations, called messages, are exchanged along graph edges. 
If the graph contains no loops, then these marginal \acp{pdf}/\acp{pmf} are exact. 
Otherwise, \ac{bp} must be performed iteratively, and the marginal \acp{pdf}/\acp{pmf} found are only approximations to the true \acp{pdf}/\acp{pmf} \cite{KolFri:B09}.
The iterative \ac{bp} recursion used for the \ac{da} subproblem operates on a loopy factor graph and therefore yields approximations to the true marginal posterior \acp{pdf}/\acp{pmf}.
This recursion is guaranteed to converge to the global optimum of the corresponding convex optimization problem, and the number of iterations required to meet a convergence criterion is bounded \cite{MeyKroWilLauHlaBraWin:J18,KroMeyCroCorHlaWil:J24}. As such, these approximations can be computed very efficiently and are highly accurate.

{We present selected message computations here; the complete set of messages and further details are provided in \onlinecite{MeyKroWilLauHlaBraWin:J18}.} 
As was shown in \onlinecite{MeyKroWilLauHlaBraWin:J18}, the prediction step of the \ac{bp}-based tracking method consists of computing the messages $\alpha_k\big(\V{\underline{y}}_k^{(j)}\big) \rmv=\rmv \alpha_k(\V{\underline{x}}_k^{(j)},\underline{r}_k^{(j)})$ for $j\in\{1,\ldots,j_{k-1}\}$.  
These messages are given by 
\[
\alpha_k\big(\underline{\V{x}}_{k}^{(j)}\rmv, \underline{r}_k^{(j)}\big) = \!\rmv \sum_{r_{k-1}^{(j)} \in \{0,1\}}\int \! 
  f\big( \underline{\V{x}}_{k}^{(j)}\rmv, \underline{r}_{k}^{(j)} \big| \V{x}_{k-1}^{(j)}, r_{k-1}^{(j)}\big)   f\big(\V{x}_{k-1}^{(j)}, r_{k-1}^{(j)} \big| \V{z}_{1:k-1}\big) \, \mathrm{d}\V{x}_{k-1}^{(j)} \ist.\nn \\[-5mm] \nn
\]
Here, $f\big(\V{x}_{k-1}^{(j)}, r_{k-1}^{(j)} \big| \V{z}_{1:k-1}\big)$ is the marginal posterior \ac{pdf} of legacy \ac{pt} $j \in \{1,\ldots,j_{k-1}\}$ at time $k-1$.
The prediction message is normalized, and we denote its predicted nonexistence probability by $\alpha_k^{(j)} \triangleq \int \alpha_k\big(\underline{\V{x}}_k^{(j)},0\big)\,\mathrm{d}\underline{\V{x}}_k^{(j)}$. For each legacy \ac{pt}, the prediction message is then combined with the current measurement model in a measurement-evaluation step. The resulting message, corresponding to Eq.~(78) in Ref.~\onlinecite{MeyKroWilLauHlaBraWin:J18}\vspace{1mm}, is
\begin{equation}
\label{eq:betaInput}
\begin{aligned}
\beta_k^{(j)}\big(a_k^{(j)}\big)
&= \sum_{\underline{r}_k^{(j)}\in\{0,1\}} \int
q\big(\underline{\V{x}}_k^{(j)},\underline{r}_k^{(j)},a_k^{(j)};\V{z}_k\big) \\[-1mm]
&\hspace{18mm}\times \alpha_k\big(\underline{\V{x}}_k^{(j)},\underline{r}_k^{(j)}\big)\,
\mathrm{d}\underline{\V{x}}_k^{(j)} \\[0.5mm]
&= \int q\big(\underline{\V{x}}_k^{(j)},1,a_k^{(j)};\V{z}_k\big)
\alpha_k\big(\underline{\V{x}}_k^{(j)},1\big)\,
\mathrm{d}\underline{\V{x}}_k^{(j)}
+1\big(a_k^{(j)}\big)\alpha_k^{(j)}\ist.
\end{aligned}
\vspace{1mm}
\end{equation}
The second equality uses $q\big(\underline{\V{x}}_k^{(j)},0,a_k^{(j)};\V{z}_k\big)=1(a_k^{(j)})$, where $1(a_k^{(j)})$ is the indicator function introduced in the definition of $q(\cdot)$ following \eqref{eq:jointPosteriorPDF}. The collection of messages $\beta_k^{(j)}(a)$, $j\in\{1,\ldots,j_{k-1}\}$, provides the inputs to \ac{pda}, whose corresponding output messages are $\kappa_k^{(j)}(a_k^{(j)})$. \Ac{pda} can be performed by brute-force marginalization \cite{BarLi:95}, MCMC techniques \cite{Del:T01}, or loopy BP \cite{WilLau:J14}.

The subsequent measurement update consists of computing the messages $\gamma_k^{(j)}\big(\underline{\V{x}}_k^{(j)}\rmv,\underline{r}_k^{(j)}\big)$.
These messages can be computed as
\begin{align*}
&\gamma_k^{(j)}\big(\underline{\V{x}}_k^{(j)}\rmv,\underline{r}_k^{(j)}\big) = \!\rmv \sum^{m_k}_{{a}^{(j)}_{k} = 0}\rmv\rmv q\big( \underline{\V{x}}^{(j)}_{k}\rmv, \underline{r}^{(j)}_{k}\rmv, a_{k}^{(j)};\V{z}_{k} \big) 
\ist \kappa_k^{(j)}\big(a_{k}^{(j)}\big).
\end{align*}
The complete \ac{bp}-based \ac{da} message computations are provided in \onlinecite{MeyKroWilLauHlaBraWin:J18,KroMeyCroCorHlaWil:J24}.
The final outputs of this message-passing algorithm are the so-called beliefs, which are highly accurate approximations of the current-time marginal posterior \acp{pdf} $f\big(\underline{\V{x}}_k^{(j)},\underline{r}_k^{(j)}\big|\V{z}_{1:k}\big)$. These marginal posterior \acp{pdf} can then be used for state estimation according to \eqref{eq:pePMF}--\eqref{eq:condPost}.

\section{Track Smoothing and Stitching}\label{sec:smoothStitch}
After applying the \ac{bp}-based \ac{mtt} approach to the beaked whale tracking problem, we obtain target trajectories, or state estimates linked across multiple time steps that correspond to the same target/whale. 
{The statistical model underlying \ac{mtt} assumes that target-measurement generation is a Bernoulli process and is therefore independent across time steps. As a result, if a target is not detected for multiple consecutive steps, the current track is terminated, and a new track is initialized if detections resume.} 
{Because of the complexity of the underwater environment and whale behavior, click detections are neither independent nor identically distributed across time steps. Long sequences of missed detections commonly occur when whales are not vocalizing or when the highly directional clicks cannot be detected above the noise. Consequently, the target-state trajectories produced by \ac{mtt} methods alone are highly fragmented and require additional processing to form a continuous track for each target. In this section, we present track smoothing and track stitching methods that improve track continuity and accuracy in the presence of missing detections.}

\subsection{Track Smoothing}

{Track smoothing improves target-state estimation at each time step $k$ by using measurements from both past and future time steps \cite{And:79,BeaVoV0:C16}. Because the original state estimate $\hat{\V{x}}_k^{(j)}$\rmv, $j \in \{1,\ldots,j_{k}\}$, is obtained using only measurements up to time $k$, incorporating future measurements can increase its accuracy.}
{To perform track smoothing, we compute marginal posterior \acp{pdf} forward and backward in time using the statistical model presented in Section~\ref{sec:statMod} and the appropriate past and future measurements. Because time is reversed for the backward computation, the state-transition model becomes}
\vspace{1mm}
\begin{equation}
\label{eq:backwardStateTransitionModel}
\V{\underline{x}}_k^{(j)} = \begin{bmatrix}
1 & 0 & t_k & 0\\
0 & 1 & 0 & t_k\\
0 & 0 & 1 & 0\\
0 & 0 & 0 & 1
\end{bmatrix}
\V{\underline{x}}_{k+1}^{(j)} - 
\begin{bmatrix}
1 & 0 & t_k & 0\\
0 & 1 & 0 & t_k\\
0 & 0 & 1 & 0\\
0 & 0 & 0 & 1
\end{bmatrix}^{-1}
\begin{bmatrix}
0.5t_k^2 & 0\\
0 & 0.5t_k^2\\
t_k & 0\\
0 & t_k\\
\end{bmatrix}\V{w}_k^{(j)}.
\end{equation}
{To maintain a consistent number of tracks in both processing directions, new tracks are introduced only during forward processing, and each backward track is initialized from the final state estimate of its corresponding forward track. This process provides a well-defined backward track for each forward track. At each time step $k$, the paired forward and backward tracks are then combined to form a smoothed track that is more accurate than either individual track.}
 
\begin{figure*}[t]
    \centering
    \hspace{10mm}
    \begin{subfigure}{0.7\textwidth}
        \begin{overpic}[width=1\textwidth,keepaspectratio]{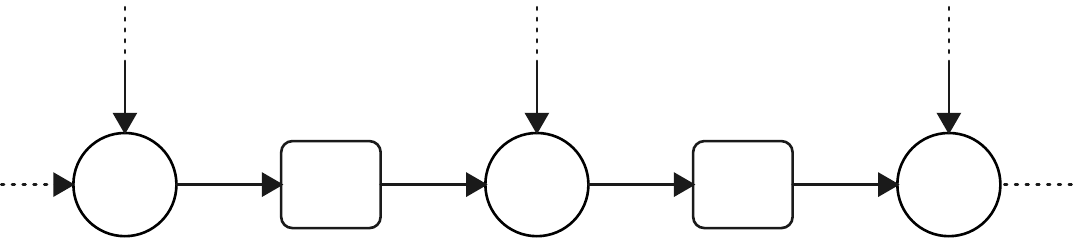}
            \put(-30,15){Forward Message}
            \put(-20,10){Passing}
            \put(8,4){$\V{x}_{k-1}^{\mathrm{f}\hspace{.05mm}(j)}$}
            \put(29,4){$f_k$}
            \put(47,4){$\V{x}_k^{\mathrm{f}\hspace{.05mm}(j)}$}
            \put(66,4){$f_{k+1}$}
            \put(85,4){$\V{x}_{k+1}^{\mathrm{f}\hspace{.05mm}(j)}$}
            \put(52,18){$\gamma_{k}^{\mathrm{f}\hspace{.05mm}(j)}$}
            \put(36,9){$\alpha_{1:k-1}^{\mathrm{f}\hspace{.05mm}(j)}$}
        \end{overpic}
    \end{subfigure}
    \vspace{5mm}
    
    \hspace{10mm}
    \begin{subfigure}{0.7\textwidth}
        \begin{overpic}[width=1\textwidth,keepaspectratio]{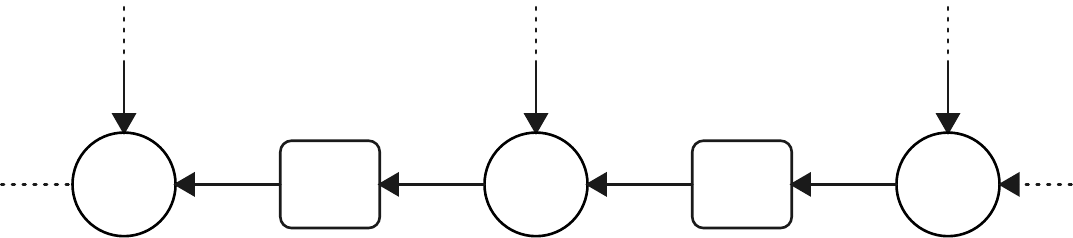}
            \put(-30,15){Backward Message}
            \put(-20,10){Passing}
            \put(8,4){$\V{x}_{k-1}^{\mathrm{b}\hspace{.05mm}(j)}$}
            \put(29,4){$f_k$}
            \put(47,4){$\V{x}_k^{\mathrm{b}\hspace{.05mm}(j)}$}
            \put(66,4){$f_{k+1}$}
            \put(85,4){$\V{x}_{k+1}^{\mathrm{b}\hspace{.05mm}(j)}$}
            \put(52,18){$\gamma_{k}^{\mathrm{b}\hspace{.05mm}(j)}$}
            \put(54,9){$\alpha_{k+1:K}^{\mathrm{b}\hspace{.05mm}(j)}$}
        \end{overpic}
    \end{subfigure}
    \vspace{5mm}
    
    \hspace{10mm}
    \begin{subfigure}{0.7\textwidth}
        \begin{overpic}[width=1\textwidth,keepaspectratio]{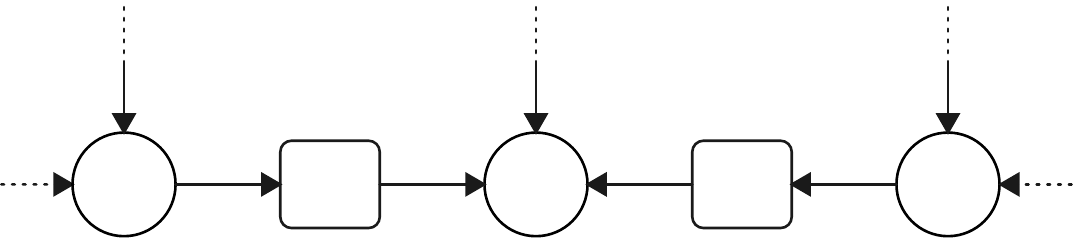}
            \put(-18,12){Smoothing}
            \put(8,4){$\V{\hat{x}}_{k-1}^{(j)}$}
            \put(29,4){$f_k$}
            \put(47,4){$\V{\hat{x}}_k^{(j)}$}
            \put(66,4){$f_{k+1}$}
            \put(85,4){$\V{\hat{x}}_{k+1}^{(j)}$}
            \put(52,18){$\gamma_{k}^{\mathrm{f}\hspace{.05mm}(j)}$}
            \put(36,9){$\alpha_{1:k-1}^{\mathrm{f}\hspace{.05mm}(j)}$}
            \put(54,9){$\alpha_{k+1:K}^{\mathrm{b}\hspace{.05mm}(j)}$}
        \end{overpic}
    \end{subfigure}
    \caption{{Diagram showing the origin of the messages used to obtain the smoothed result. We use the shorthand $f_k$, where $f_k=f(\V{x}_k|\V{x}_{k-1})$. The bottom diagram shows how the smoothed estimate is obtained from the product of the messages $\gamma_{k}^{\mathrm{f}\hspace{.05mm}(j)}$, $\alpha_{1:k-1}^{\mathrm{f}\hspace{.05mm}(j)}$, and $\alpha_{k+1:K}^{\mathrm{b}\hspace{.05mm}(j)}$, thereby incorporating information from both $1:k$ and $k+1:K$, where $K$ is the index of the final time step.}}
    \label{fig:smoothing}
\end{figure*}

{The approximate marginal posterior \acp{pdf} used for track smoothing can also be computed by \ac{bp} message passing. This process, shown in Fig.~\ref{fig:smoothing}, uses the same factor graph conventions as the factorization of the joint posterior \ac{pdf} in \ac{bp}-based \ac{mtt}. According to the \ac{bp} message-passing rules \cite{KolFri:B09}, the messages $\alpha_k^{\mathrm{f}\hspace{.05mm}(j)}(\V{x}_k^{(j)}\rmv,r_k^{(j)})$ and $\gamma_k^{(j)}(\V{x}_k^{(j)}\rmv,r_k^{(j)})$ are passed forward in time, and the message $\alpha_k^{\mathrm{b}\hspace{.05mm}(j)}(\V{x}_k^{(j)}\rmv,r_k^{(j)})$ is passed backward in time. The approximate marginal posterior \ac{pdf} $\tilde{f}(\V{x}_k^{(j)}\rmv,r_k^{(j)}|\V{z}_{1:K})$ is then given by}
\begin{equation}
\label{eq:smooth}
\tilde{f}\big(\V{x}_k^{(j)}\rmv,r_k^{(j)}|\V{z}_{1:K}\big) \propto \alpha_k^{\mathrm{f}\hspace{.05mm}(j)}\big(\V{x}_k^{(j)}\rmv,r_k^{(j)}\big)\ist\gamma_k^{(j)}\big(\V{x}_k^{(j)}\rmv,r_k^{(j)})\ist\alpha_k^{\mathrm{b}\hspace{.05mm}(j)}\big(\V{x}_k^{(j)}\rmv,r_k^{(j)}\big)\ist. 
\end{equation}

\subsection{Track Stitching}
{Another strategy for improving target-state estimation, particularly when tracks are fragmented, is track stitching \cite{RagSriThaKir:J18}. The goal of track stitching is to determine which track fragments are associated and thereby identify the continuous target track to which each group of fragments belongs. Although association can be performed on all fragments simultaneously, we reduce its complexity by using a sliding window of length $B$ seconds. Within each window, all track fragments that overlap the window are divided into old tracks $\hat{\V{x}}^{(o)}\rmv\in\rmv T_{\text{O}}$ and new tracks $\hat{\V{x}}^{(n)} \in T_{\text{N}}$. If the final state estimate of a track lies within the window, the track is classified as old; otherwise, it is classified as new. As in the conventional \ac{da} problem, in which the relationships between measurements and targets are unknown, track association determines which old track $\hat{\V{x}}^{(o)}\in T_{\text{O}}$ corresponds to which new track $\hat{\V{x}}^{(n)} \in T_{\text{N}}$. We formulate track association as a cost-minimization problem, assign a cost to every possible pairing of old and new tracks, and identify the joint association with the lowest cost. Costs are also assigned to ``dummy pairings,'' in which an old or new track is not associated with a corresponding track.} 

The construction below can be viewed as a temporal track-to-track association test \cite{BarLi:95,BarWilTia:B11,TiaBar:J09,Gov:B23}. As in classical track-to-track association, two state estimates are expressed at a common time, their covariance-weighted difference is used to evaluate whether they have a common origin, and a one-to-one assignment is selected. There are, however, two important distinctions. First, the sliding window test in \onlinecite{TiaBar:J09} aggregates synchronized estimates from continuing sensor tracks, whereas our window limits the set of candidate fragments and our cost compares one propagated endpoint with one new-fragment state. Second, classical track-to-track fusion subsequently combines contemporaneous estimates; here, a new fragment that starts before the old fragment ends is inadmissible, and accepted fragments are concatenated, so no state-estimate fusion is performed at the join.

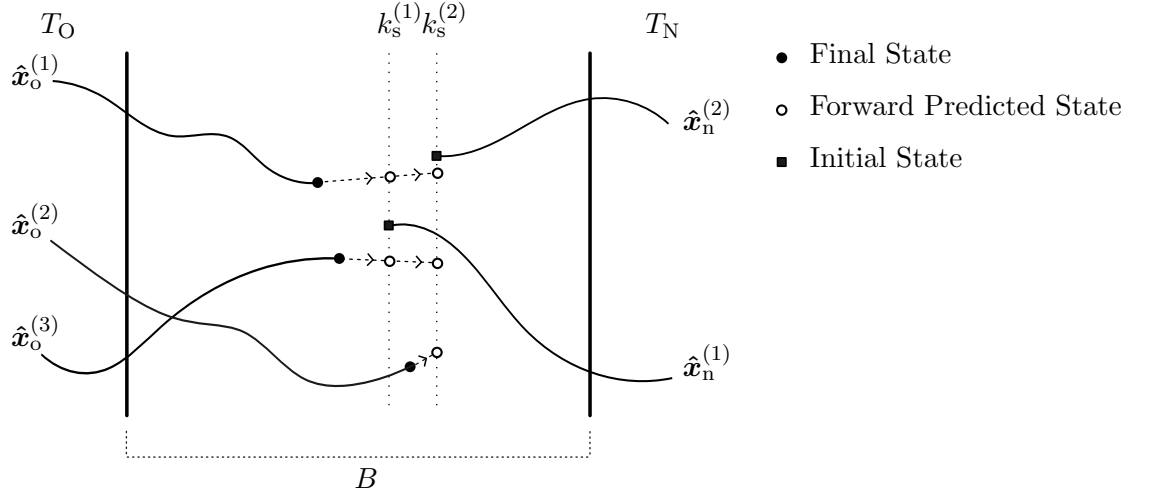
\begin{figure*}[t]
    \vspace{5mm}
    \centering
    \hspace{-10mm}
        \begin{overpic}[width=0.6\textwidth,keepaspectratio]{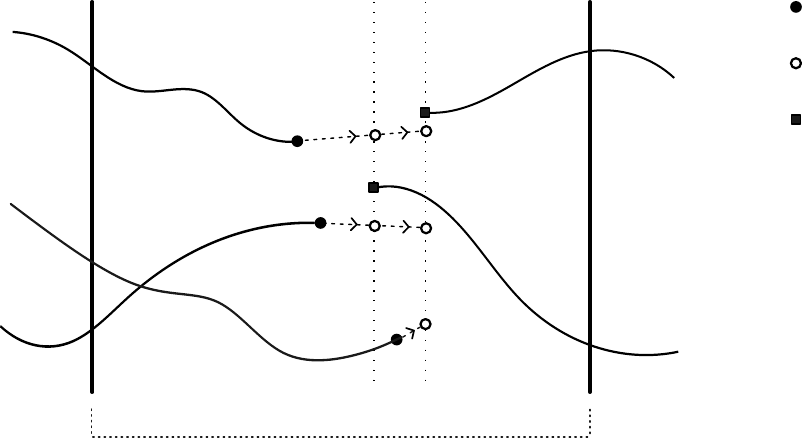}
            \put(0,57){\small$T_\text{O}$}
            \put(81,57){\small$T_{\text{N}}$}
            \put(45,57){\small$k_{\text{s}}^{(1)}$}
            \put(51,57){\small$k_{\text{s}}^{(2)}$}
            \put(42,-4){\small$B$}
            \put(-4,15){\small$\V{\hat{x}}_{\text{o}}^{(3)}$}
            \put(-4,30){\small$\V{\hat{x}}_{\text{o}}^{(2)}$}
            \put(-4,50){\small$\V{\hat{x}}_{\text{o}}^{(1)}$}
            \put(86,11){\small$\V{\hat{x}}_{\text{n}}^{(1)}$}
            \put(86,44){\small$\V{\hat{x}}_{\text{n}}^{(2)}$}
            \put(103,53){\small {Final State}}
            \put(103,46){\small Forward Predicted State}
            \put(103,39){\small {Initial State}}
        \end{overpic}
    \vspace{8mm}
    \caption{{Visualization\cite{MaKroBagBauMey:C25} of the track stitching process for a window with a duration of $B$ seconds. The final state of each track in the set of old tracks, $T_\text{O}$, is predicted to each time step {$k_{\text{s}}^{(n)}$} corresponding to an initial state in the set of new tracks, $T_\text{N}$.}}
    \label{fig:stitching}
    \vspace{-1mm}
\end{figure*}

We set up the track association as follows: 
{For a given pairing of an old track and a new track $\big(\V{\hat{x}}^{(o)}\rmv,\V{\hat{x}}^{(n)}\big)$, let $k_{\text{e}}^{(o)}$ denote the final time step of the old track $\V{\hat{x}}^{(o)}$ and $k_{\text{s}}^{(n)}$ the initial time step of the new track $\V{\hat{x}}^{(n)}$.} 
{Using the final state estimate of $\V{\hat{x}}^{(o)}\rmv$ at time step $k_{\text{e}}^{(o)}$, we predict the old track forward in time under the state-transition model described in Section~\ref{sec:statMod}. The resulting predicted state estimate at time step $k_{\text{s}}^{(n)}$ is denoted by $\hat{\V{x}}_{k_{\text{s}}^{(n)}|k_{\text{e}}^{(o)}}^{(o)}$. This process is visualized in Fig.~\ref{fig:stitching}, where each of the three old tracks in $T_{\text{O}}$ is predicted forward to the initial time step of each of the two new tracks in $T_{\text{N}}$.}
{Next, we compare $\hat{\V{x}}_{k_{\text{s}}^{(n)}|k_{\text{e}}^{(o)}}^{(o)}$ with the first state estimate of the new track $\V{\hat{x}}^{(n)}$ at its initial time $k_{\text{s}}^{(n)}$, denoted by $\V{\hat{x}}_{k_{\text{s}}^{(n)}}^{(n)}$, and compute their difference as} 
\begin{equation}
\label{eq:stitchDiff}
\Delta^{(o,n)} = \hat{\V{x}}_{k_{\text{s}}^{(n)}|k_{\text{e}}^{(o)}}^{(o)} - \V{\hat{x}}_{k_{\text{s}}^{(n)}}^{(n)}\ist.
\end{equation}
The covariance matrices corresponding to $\hat{\V{x}}_{k_{\text{s}}^{(n)}|k_{\text{e}}^{(o)}}^{(o)} $ and $\V{\hat{x}}_{k_{\text{s}}^{(n)}}^{(n)}$ are denoted by $\Sigma_{k_{\text{s}}^{(n)}|k_{\text{e}}^{(o)}}^{(o)}$ and {$\Sigma_{k_{\text{s}}^{(n)}|k_{\text{s}}^{(n)}}^{(n)}$}, respectively, so the accumulated covariance matrix corresponding to $\Delta^{(o,n)}$ is defined as 
\begin{equation}
\label{eq:covDiff}
\Sigma^{(o,n)} = \Sigma_{k_{\text{s}}^{(n)}|k_{\text{e}}^{(o)}}^{(o)} + {\Sigma_{k_{\text{s}}^{(n)}|k_{\text{s}}^{(n)}}^{(n)}}.
\end{equation}
Equation~\eqref{eq:covDiff} is the independent-error special case of classical track-to-track association. If the estimation errors of the two fragments have cross-covariance $\Sigma_{\times}^{(o,n)}$, the covariance of their difference also contains the terms $-\Sigma_{\times}^{(o,n)}-\big(\Sigma_{\times}^{(o,n)}\big)^{\mathsf T}$ \cite{Bar:J81,TiaBar:J09}. Cross-covariance is central to conventional track-to-track fusion, particularly when local tracks contain common process noise or previously shared information \cite{GovKoc:J12}. Treating the fragments as independent is consistent with their representation as separate potential targets before stitching; under the common-origin hypothesis, however, a gap-spanning joint model could introduce residual dependence through common motion uncertainty. Estimating this dependence, or extending the association cost to use multiple states near the fragment endpoints, is a potential direction for improving track stitching. As is, we find the association costs using the single states at the endpoints.
{If $k_{\text{e}}^{(o)} > k_{\text{s}}^{(n)}$, i.e., if the old track $\hat{\V{x}}^{(o)}$ ends after the new track $\V{\hat{x}}^{(n)}$ starts, we set the cost $c^{(o,n)}$ of that pairing to $\infty$. Otherwise, using \eqref{eq:stitchDiff} and \eqref{eq:covDiff}, we define the association cost for $\hat{\V{x}}^{(o)}$ and $\V{\hat{x}}^{(n)}$ as follows \cite{RagSriThaKir:J18}:}   
\vspace{0.5mm} 
\begin{equation}
\label{eq:stitchingCost}
c^{(o,n)}=
  \begin{cases}
  -\ln{\frac{p_{\text{T}}\ist \mathcal{N}(\Delta^{(o,n)}; 0,\Sigma^{(o,n)})}{\mu_{\text{b}} }}, & \text{for $o\ne0, n\ne0$}\\[0.5mm]
  -\ln{(1-p_{\text{T}})}, & \text{for $o=0$ or $n=0$}.
  \end{cases} 
\end{equation}
{Here, $p_{\text{T}}$ represents the probability that the old track survives to the initial time step of the new track, and $\mu_{\text{b}}$ represents the mean spatial density of newly born tracks. A ``dummy pairing'' occurs when either $o=0$ or $n=0$, where $\hat{\V{x}}^{(0)}$ represents a ``dummy track'' and the non-dummy track remains unassociated. If the non-dummy track is old, the pairing represents a true target trajectory that terminates at the end of the old track. If the non-dummy track is new, the pairing represents a true target trajectory that begins at the start of the new track.} 

{Once the costs of all possible pairs of old and new tracks within a window have been determined, the cost-minimization problem can be solved using an efficient assignment algorithm, such as the Hungarian algorithm \cite{Kuh:J55}. The lowest-cost assignment is accepted, the corresponding track fragments are merged, and the updated fragments are used in subsequent sliding window assignments. This process is repeated for every position of the sliding window. Alternatively, a greedy method can calculate the lowest-cost assignment for each window without immediately updating the tracks. The assignments from all windows are then sorted by cost, and tracks are merged in ascending order of cost; assignments that conflict with an earlier merge are discarded. On our beaked whale data, however, this method yielded no significant improvement over updating the tracks after each sliding window assignment.}

\section{Simulation Experiments}\label{sec:simulation}

{First, we demonstrate the effectiveness of our \ac{bp}-based \ac{mtt} method with track smoothing and track stitching using simulated data.}
{Tracks were generated using the constant-velocity state-transition model with additive zero-mean Gaussian noise described in Section~\ref{sec:statMod}.} 
{Next, a sensor generated noisy azimuth and elevation measurements from the target positions.}
{We simulated a two-target scenario with $300$ time steps. Each track contained $10$ measurement gaps, with each gap lasting a randomly selected duration of up to $14$ time steps. Each time step corresponded to one second.}
The gaps reflect the whales' behavior when they stop vocalizing or turn away from the receivers. 

We {used} a particle-based implementation of our \ac{bp}-based \ac{mtt}. The simulation parameters were chosen as follows:
{The driving noise variance was set to $\sigma_\mathrm{w}^2=0.0004$, and the azimuth and elevation measurement noise variances were set to $\sigma_\mathrm{v,a}^2=\sigma_\mathrm{v,e}^2=1$. The detection probability was set to $p_\mathrm{d}=0.9$, and the mean number of clutter measurements per time step was set to $\mu_\mathrm{c}=2$.}  
We used a target detection threshold of $\gamma_\mathrm{D}=0.5$. 
Track stitching used a window size of $B=50$ time steps, and the probability of stitching was set to $p_\mathrm{T}=0.95$.

\begin{figure*}[t]
    \centering
    \begin{subfigure}{0.5\textwidth}
        \includegraphics[width=\linewidth, clip]{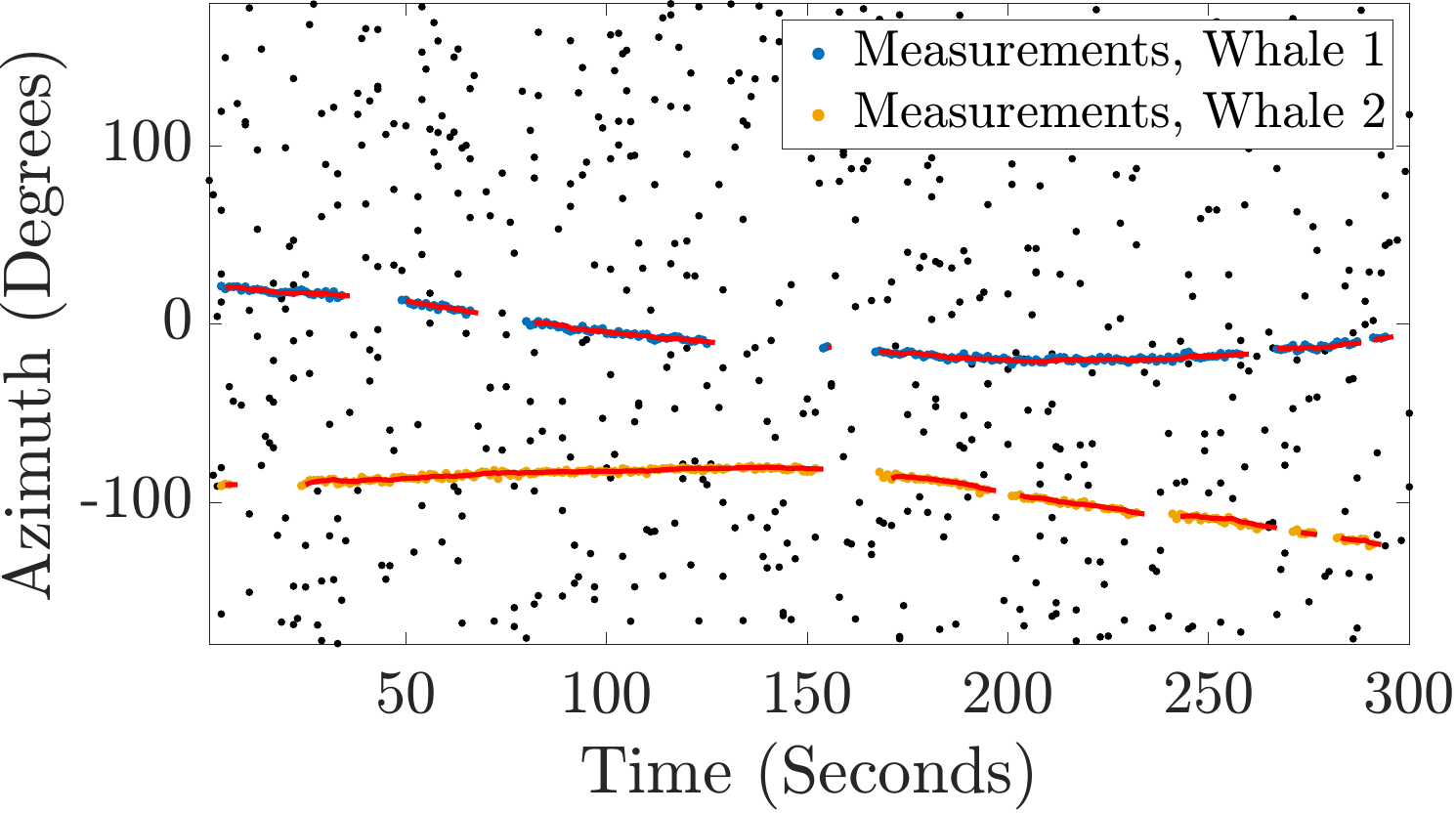}
        \subcaption{}
    \end{subfigure}
        ~ 
    \begin{subfigure}{0.5\textwidth}
        \includegraphics[width=\linewidth, clip]{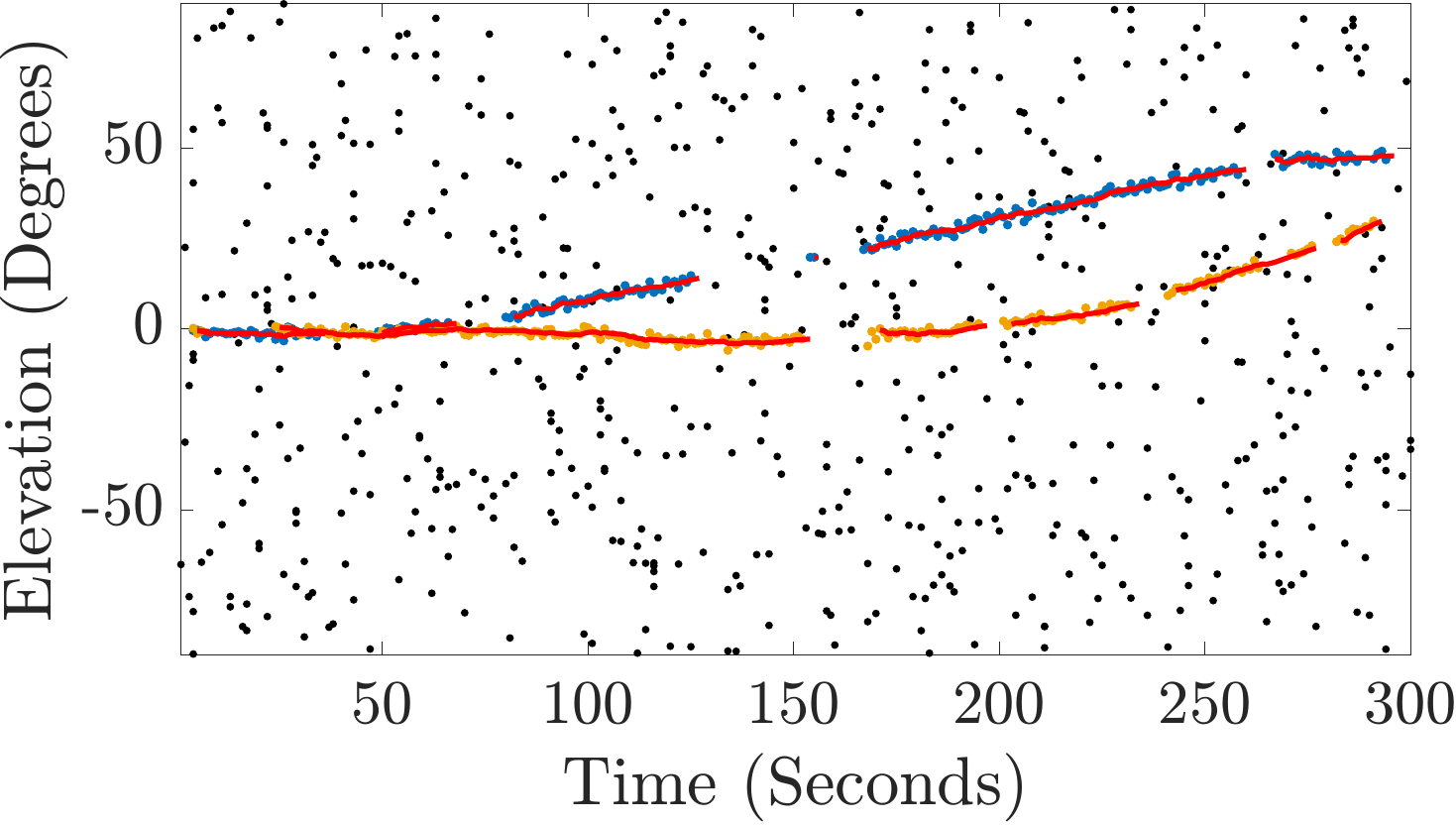}
        \subcaption{}
    \end{subfigure}

    \begin{subfigure}{0.5\textwidth}
        \includegraphics[width=\linewidth, clip]{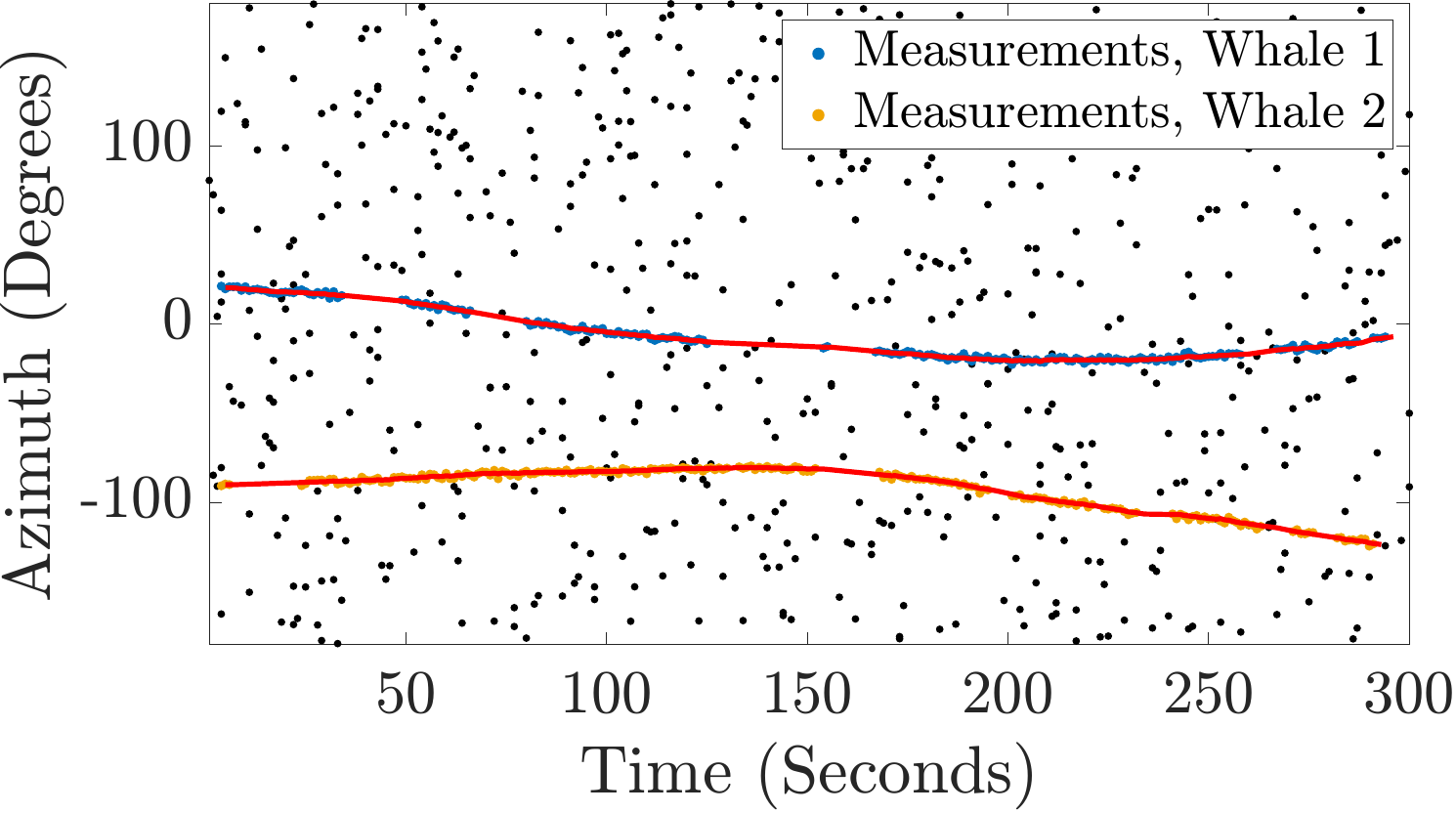}
        \subcaption{}
    \end{subfigure}
        ~ 
    \begin{subfigure}{0.5\textwidth}
        \includegraphics[width=\linewidth, clip]{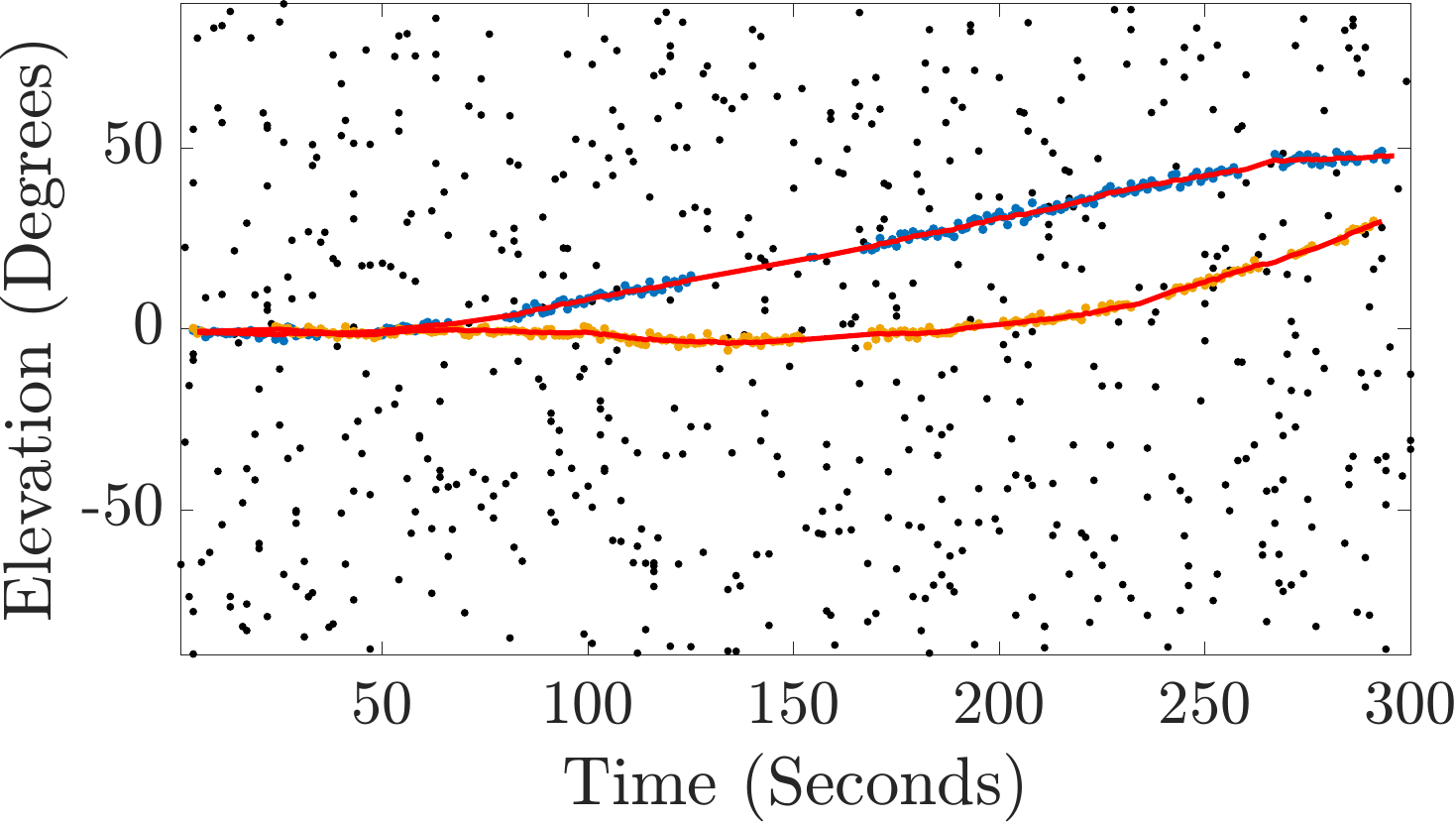}
        \subcaption{}
    \end{subfigure}
    
    \caption{{{Azimuth and elevation tracking results for two simulated tracks with randomized detection gaps. Results were obtained using \ac{bp}-based \ac{mtt} before track smoothing and stitching (a, b) and after track smoothing and stitching (c, d).}}} 
    \label{fig:simtracks}
    \vspace{-3mm}
\end{figure*}

{We compare our proposed \ac{bp}-based \ac{mtt} method with the \ac{gmphd} method, which has previously been applied to tracking false killer whales using \ac{tdoa} measurements \cite{GruNosOle:J21}.} 
{The \ac{phd} is a function whose peaks indicate the most probable target positions and whose integral over the state space equals the expected number of targets present.}
{The \ac{gmphd} approximates the \ac{phd} using a Gaussian mixture, in which the Kalman filter predicts and updates the means and covariances of the Gaussian components over time, while the \ac{phd} filter equations predict and update the Gaussian weights.} 
In contrast to our \ac{bp} approach, the \ac{gmphd} method does not perform \ac{pda}, thereby reducing computational complexity at the cost of potentially lower tracking performance \cite{KroMeyCorCarMenWil:J21}. 
For our comparison, we used a maximum {of} $100$ Gaussian components, a Mahalanobis distance threshold of four for component merging, a component pruning threshold of $0.001$, and a $0.1$ weight threshold for state estimation. 
{The driving noise variance was set to $\sigma_\mathrm{w}^2=0.01$, and the mean number of new targets was set to $\mu_\mathrm{b}=0.0001$. Otherwise, the parameters were the same as those used for the \ac{bp}-based \ac{mtt}.}

\begin{figure*}[t]
    \centering
    \begin{subfigure}{1\textwidth}
        \includegraphics[width=\linewidth, clip]{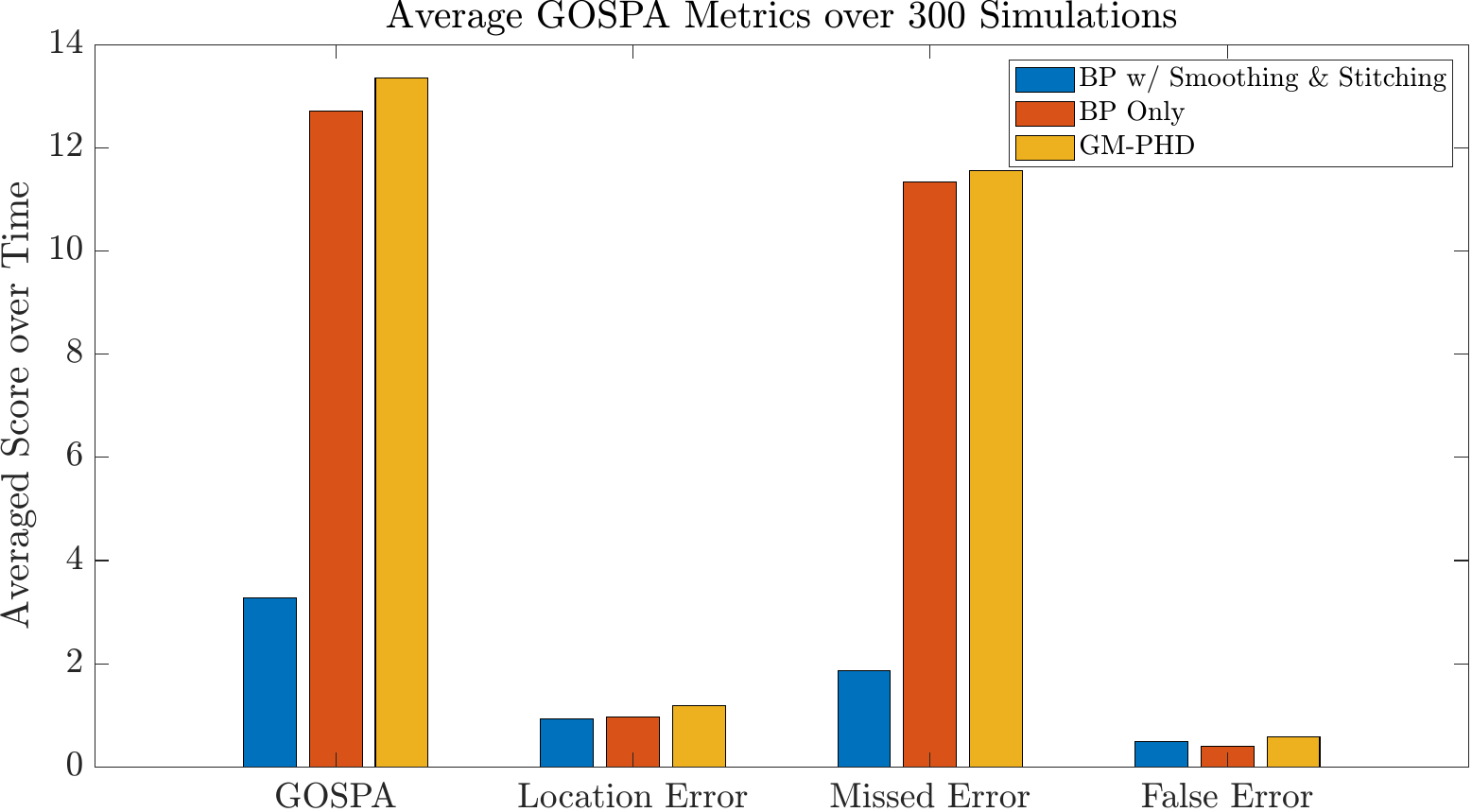}
        \subcaption{}
    \end{subfigure}%
    
    \caption{{\Ac{gospa} metrics averaged over 300 simulated scenarios of two tracks with randomized measurement detection gaps.}} 
    \label{fig:simgospa}
    \vspace{-3mm}
\end{figure*}

{Fig.~\ref{fig:simtracks} shows a single simulation and compares results obtained using our proposed \ac{bp}-based \ac{mtt} method without and with track smoothing and track stitching.} 
{As expected, our \ac{bp}-based method accurately follows continuous sequences of detections but, without track stitching, quickly terminates tracks once a target is no longer detected.} 
Using track stitching, the track segments were correctly linked together.
{Fig.~\ref{fig:simgospa} shows the \ac{gospa} metric averaged over all $300$ time steps and $300$ Monte Carlo runs. We compare our \ac{bp}-based \ac{mtt} method with and without track smoothing and stitching against the \ac{gmphd} method used in \onlinecite{GruNosOle:J21}.}
{The missed-target error is significantly lower for \ac{bp}-based \ac{mtt} with track smoothing and stitching than for either \ac{bp}-based \ac{mtt} without these processing steps or the \ac{gmphd} filter.} 
{The remaining error components differ only slightly among the methods; consequently, our proposed method achieves a significantly lower total \ac{gospa} error than the reference methods.} 

\section{\texorpdfstring{{Real-World Application}}{Real-World Application}}\label{sec:results}
Next, we evaluate our proposed \ac{bp}-based \ac{mtt} method with track smoothing and track stitching using acoustic recordings of \textit{Ziphius cavirostris} collected from March to May 2022 at a site south of Santa Rosa Island ($33^{\circ}32.221'~\mathrm{N}$, $120^{\circ}14.856'~\mathrm{W}$) in the Southern California Bight \cite{BagSnyBerCurWigHilSirFraBau:J25}. 
Data were collected simultaneously using two \acp{harp} \cite{WigHil:C07} deployed on the seafloor. 
Each \ac{harp} was equipped with a small-aperture array of four hydrophones mounted $6\,\mathrm{m}$ above the seafloor and arranged tetrahedrally with approximately $1\,\mathrm{m}$ spacing. The \acp{harp} recorded continuously at a sampling rate of $100\,\mathrm{kHz}$. Individual echolocation clicks were detected and localized by expert analysts using the software package \textit{Where's Whaledo}\cite{SnySolSimFraWigHil:J24,BagSnyBerCurWigHilSirFraBau:J25}. Clicks were represented first as \acp{tdoa} between the respective hydrophone pairs, which were used to estimate the \ac{doa}, represented by azimuth and elevation angles \cite{BagSnyBerCurWigHilSirFraBau:J25}. These \ac{doa} click detections were used as the measurements in our \ac{mtt} method.
Each \ac{harp} was equipped with a small-aperture array of four hydrophones mounted $6\,\mathrm{m}$ above the seafloor and arranged tetrahedrally with approximately $1\,\mathrm{m}$ spacing. The \acp{harp} recorded continuously at a sampling rate of $100\,\mathrm{kHz}$. Individual echolocation clicks were detected and localized by expert analysts using the software package \textit{Where's Whaledo}\cite{SnySolSimFraWigHil:J24,BagSnyBerCurWigHilSirFraBau:J25}. Clicks were represented first as \acp{tdoa} between the respective hydrophone pairs, which were used to estimate the \ac{doa}, represented by azimuth and elevation angles \cite{BagSnyBerCurWigHilSirFraBau:J25}. These \ac{doa} click detections were used as the measurements in our \ac{mtt} method.

{Although \ac{mtt} is well suited to and has frequently been applied in active sonar and radar settings, tracking marine mammals using passive acoustic recordings differs substantially from those applications and can therefore be more challenging. As discussed by Gruden, Nosal, and Oleson \cite{GruNosOle:J21}, $p_\mathrm{d}$ in sonar and radar depends only on the \ac{snr}, whereas $p_\mathrm{d}$ in \ac{pam} depends on both target detectability and target availability. Here, detectability is the probability of detecting a whale given that it has produced a click, whereas availability is the probability that the whale clicks. Detectability depends on the source level and the \ac{snr} but is also affected by the highly directional echolocation clicks of \textit{Ziphius cavirostris} \cite{GasWigHil:J15,ZimJohMadTya:J05}. Consequently, detectability varies over time with whale movement and is difficult to model. Furthermore, availability is dependent on where the whale is in their dive cycle, as beaked whales click mostly while diving at greater depths\cite{HilBauFraTriMerWigMcdGarHarMarTho:J15}. Even while clicking regularly, though the average \ac{ici} of \textit{Ziphius cavirostris} is known, short pauses may occur between click trains\cite{FraGooSkaTarKan:J02}, making availability difficult to estimate. This behavioral complexity produces the previously discussed gaps in click detections, which a constant $p_\mathrm{d}$ does not represent. We account for this uncertainty by using a lower $p_\mathrm{d}$ and applying track stitching during post-processing.}

{Unlike active sonar or radar systems, our passive acoustic recording system requires a time grid that aligns clicks with discrete time steps. Because the intervals between clicks are irregular, the intervals between time steps are also nonconstant. We construct the time grid by first dividing the entire encounter into $0.4$-s intervals and then assigning each time step the time of the highest-amplitude click within its interval. All other clicks within that interval are aligned with the same time step. If an interval contains no clicks, no time step is assigned. Removing intervals without measurements is necessary because of the high variability in \ac{ici} and prevents the filter from treating even short pauses in whale clicks as multiple missed detections. The initial interval of $0.4$ s approximates the lower bound on the \ac{ici} of \textit{Ziphius cavirostris} \cite{ZimJohMadTya:J05}, thereby preventing multiple clicks from the same whale from being aligned with the same time step.}  

{For the results obtained using \textit{Ziphius cavirostris} data, we used the following empirically determined parameters. The driving noise variance in the constant-velocity state-transition model was set to $\sigma_\mathrm{w}^2=2.5\times10^{-5}$. The azimuth and elevation measurement noise variances were set to $\sigma_\mathrm{v,a}^2=\sigma_\mathrm{v,e}^2=2.56$. The marginal variances in the joint velocity-covariance prior were set to $\sigma_\mathrm{p,a}^2=\sigma_\mathrm{p,e}^2=0.64$, and the azimuth and elevation velocities were assumed to be independent in this prior. The mean number of clutter detections at each step was set to $\mu_\mathrm{c}=5$, and the mean number of newly detected targets was set to $\mu_\mathrm{b}=0.002$. The detection probability was set to $p_\mathrm{d}=0.7$, and the target-existence threshold was set to $\gamma_\mathrm{D}=0.5$. Because the time grid is irregular, we used a time-dependent survival probability. We set $p_\mathrm{s}=p_\mathrm{si}=0.98$ for the base \ac{ici} of $0.4$ s. For an interval of length $t_k$, we then used the ratio $q=t_k/0.4$ to set $p_\mathrm{s}=p_\mathrm{si}^q$. The track stitching window length was set to $B=300$ s, and the stitching probability was set to $p_\mathrm{T}=0.95$. After track smoothing and stitching, tracks with fewer than 30 estimated states were assumed to be false or duplicate tracks and were removed.}

The particle-based implementation used $5000$ particles. Potential targets were pruned when their posterior existence probabilities fell below $1.2\times10^{-5}$. The additional hard gap-pruning rule was disabled by setting its time threshold to infinity; thus, a potential target was not pruned solely because the interval to the next measurement exceeded a fixed duration.

{We compare our results for real \textit{Ziphius cavirostris} data with those produced by the \ac{gmphd}. Again, we used a maximum of 100 components, a Mahalanobis-distance threshold of 4 for merging components, a weight threshold of 0.001 for pruning components, and a weight threshold of 0.1 for state estimation.} 
{The driving noise variance was set to $\sigma_\mathrm{w}^2=0.0025$. Otherwise, the parameters were the same as those used for \ac{bp}-based \ac{mtt}.} Comparisons were made by evaluating \ac{gospa} using $p=1$, cutoff $c=10$, and $\alpha=2$, with azimuth differences treated circularly by wrapping them to $[-180^\circ,180^\circ)$. Each metric was time-normalized within an encounter, averaged equally over the five encounters, and then averaged over runs with random seeds 1--10. For each seed and encounter, the random draws used for \ac{bp} filtering, smoothing, and the \ac{gmphd} were generated from separate, repeatable substreams. The potential-target pruning threshold and hard gap-pruning setting were selected using these same seeds; the resulting comparison is therefore descriptive and post-selection rather than a held-out performance estimate.

\begin{figure*}[t]
    \centering
    \begin{subfigure}{0.5\textwidth}
        \includegraphics[width=\linewidth, clip]{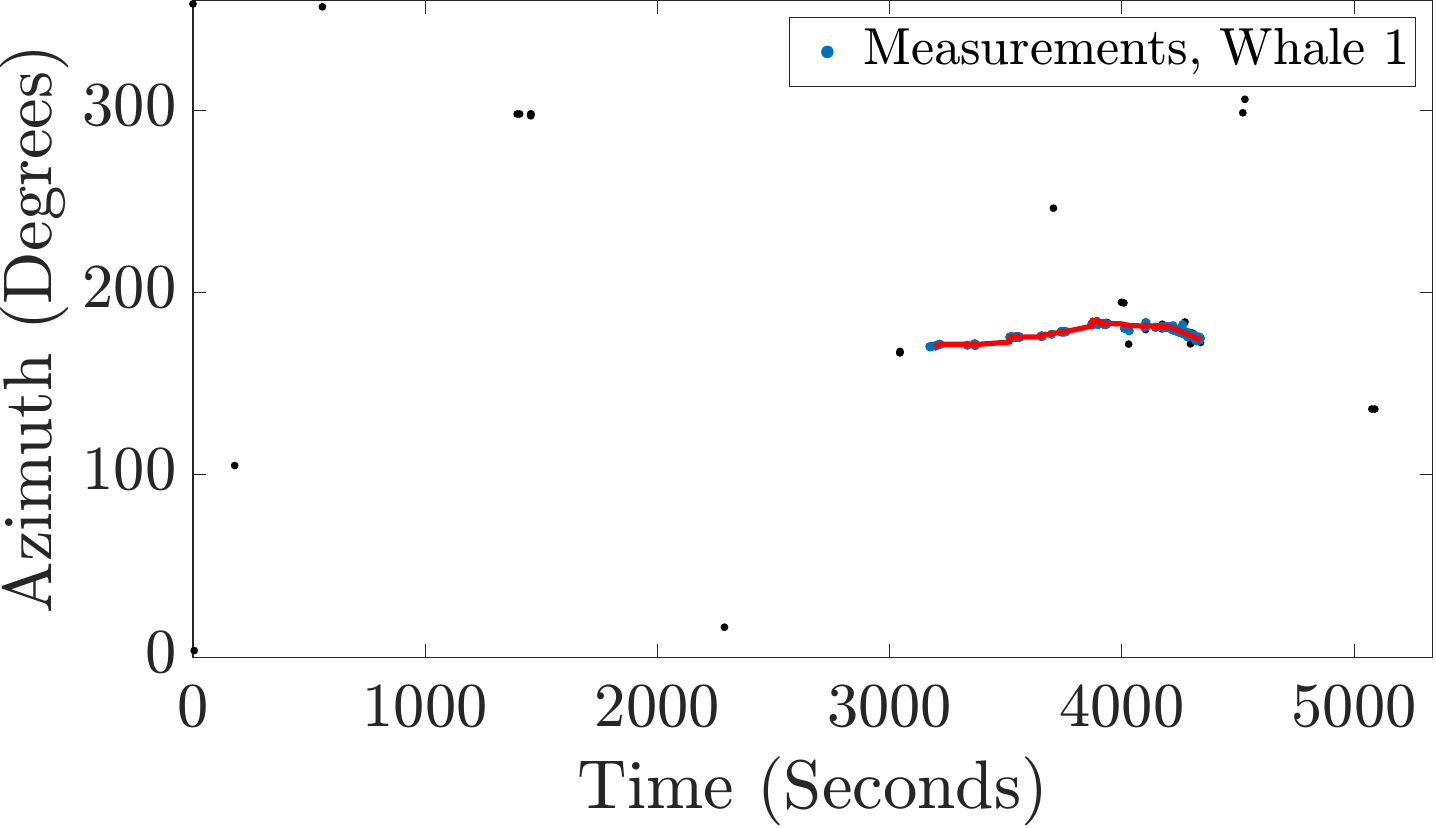}
        \subcaption{}
    \end{subfigure}%
        ~ 
    \begin{subfigure}{0.5\textwidth}
        \includegraphics[width=\linewidth, clip]{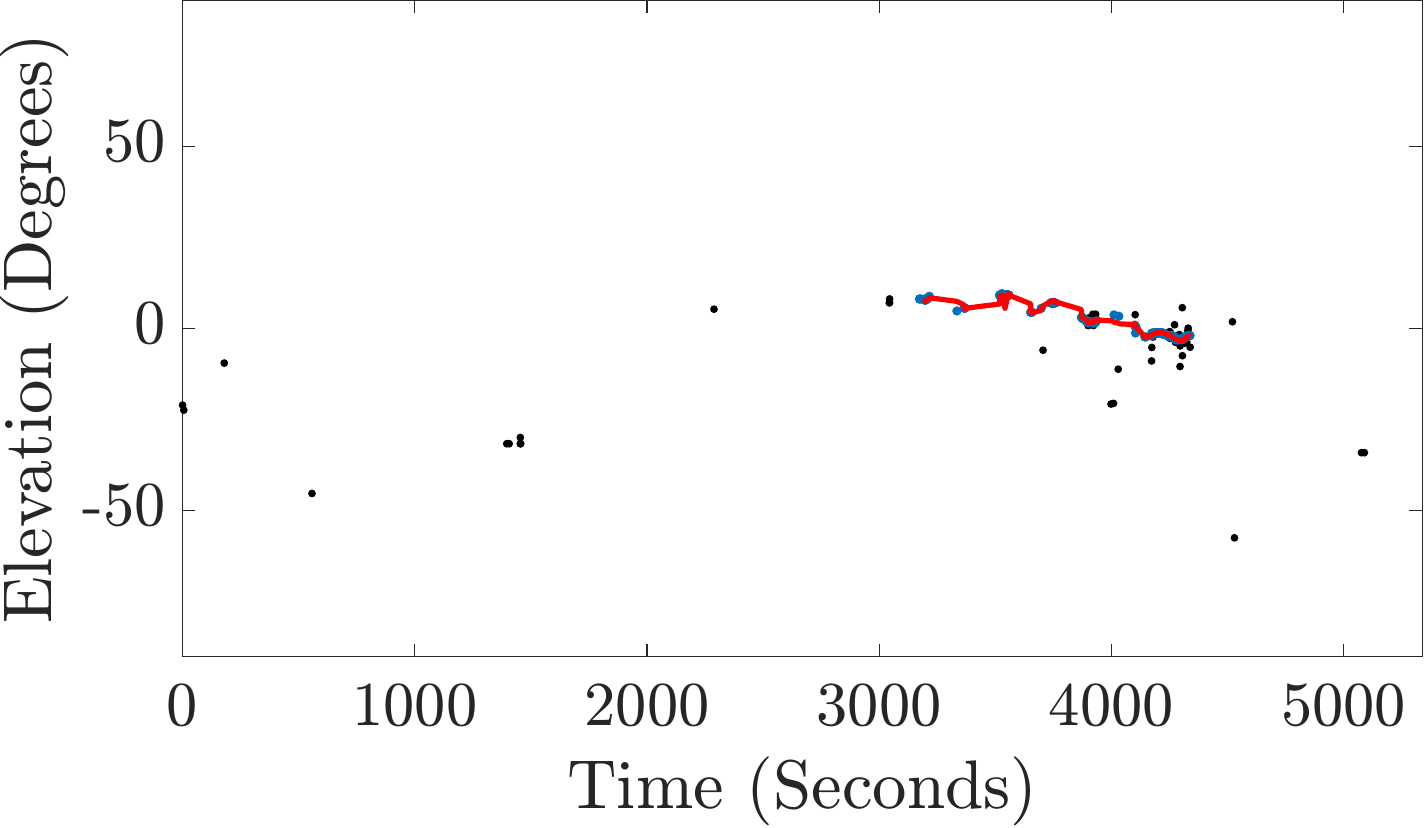}
        \subcaption{}
    \end{subfigure}

    \begin{subfigure}{0.5\textwidth}
        \includegraphics[width=\linewidth, clip]{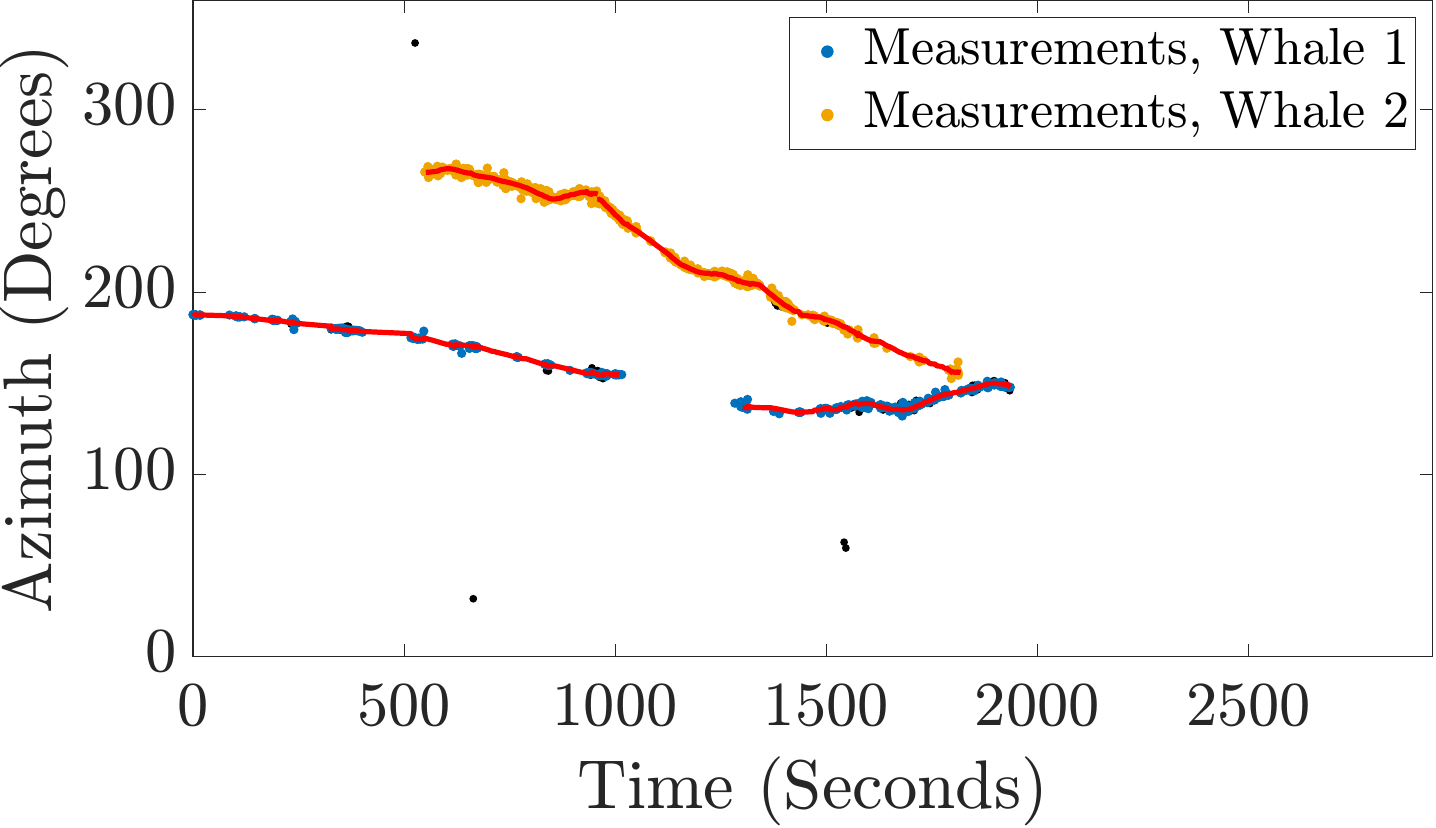}
        \subcaption{}
    \end{subfigure}%
        ~ 
    \begin{subfigure}{0.5\textwidth}
        \includegraphics[width=\linewidth, clip]{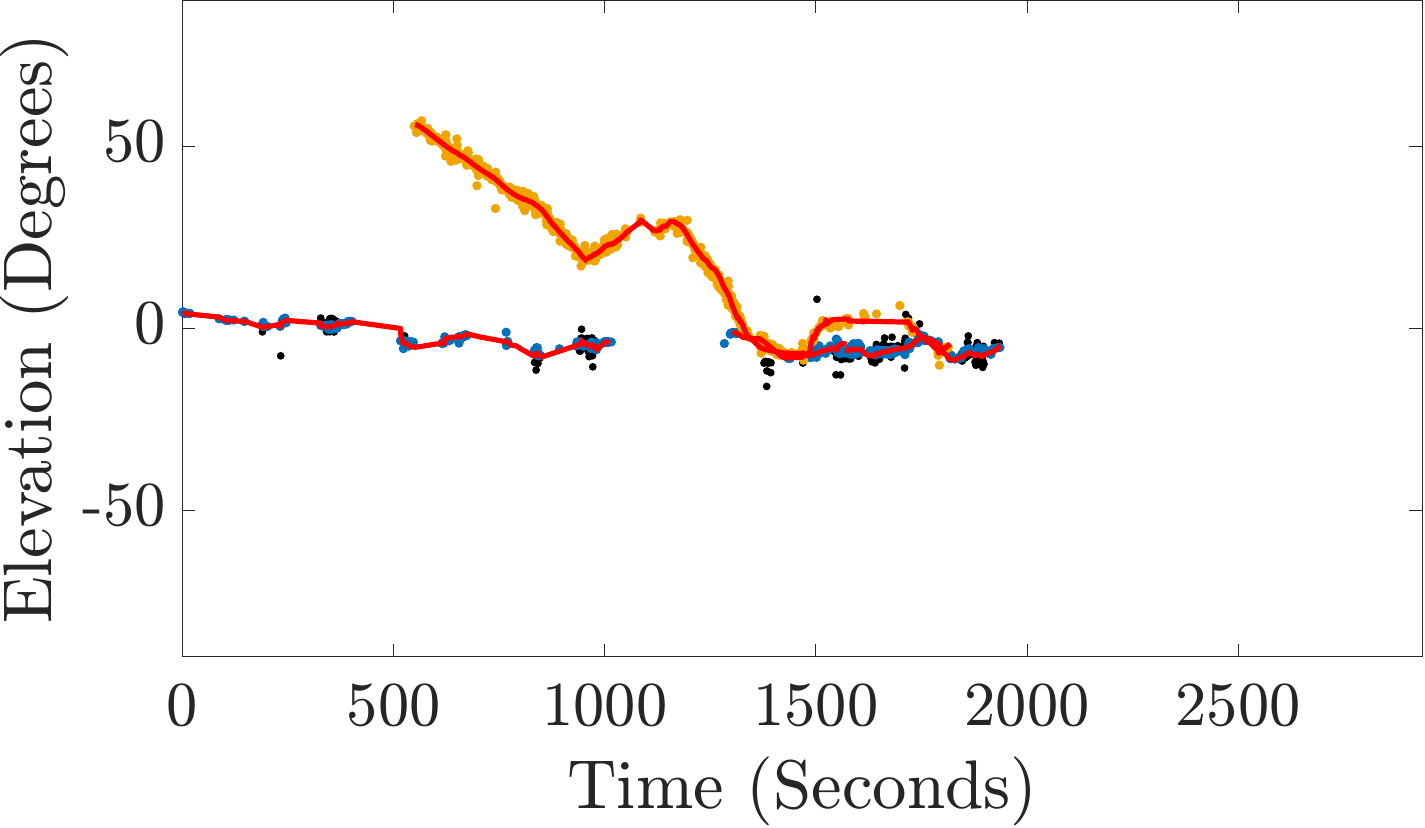}
        \subcaption{}
    \end{subfigure}

    \begin{subfigure}{0.5\textwidth}
        \includegraphics[width=\linewidth, clip]{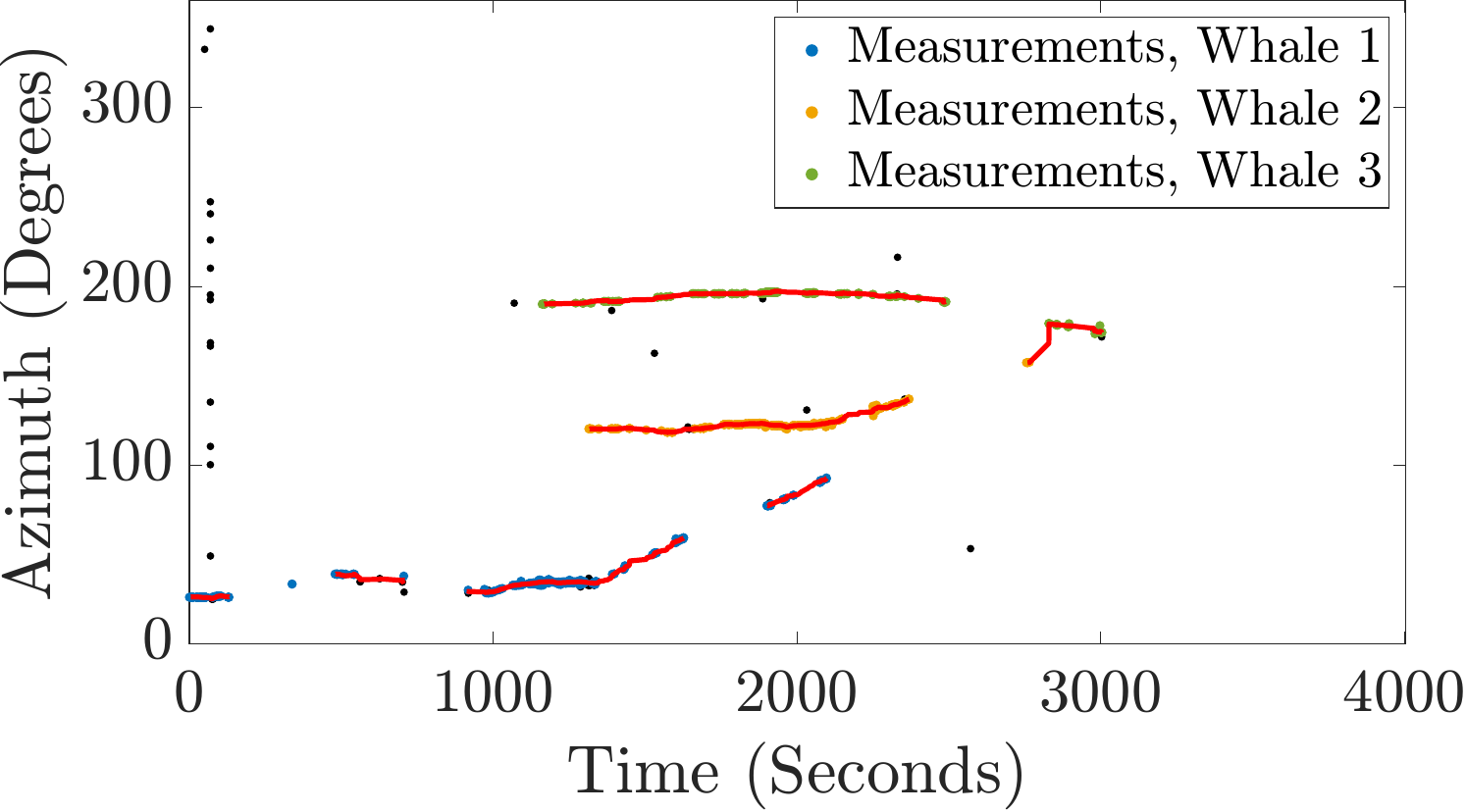}
        \subcaption{}
    \end{subfigure}%
        ~ 
    \begin{subfigure}{0.5\textwidth}
        \includegraphics[width=\linewidth, clip]{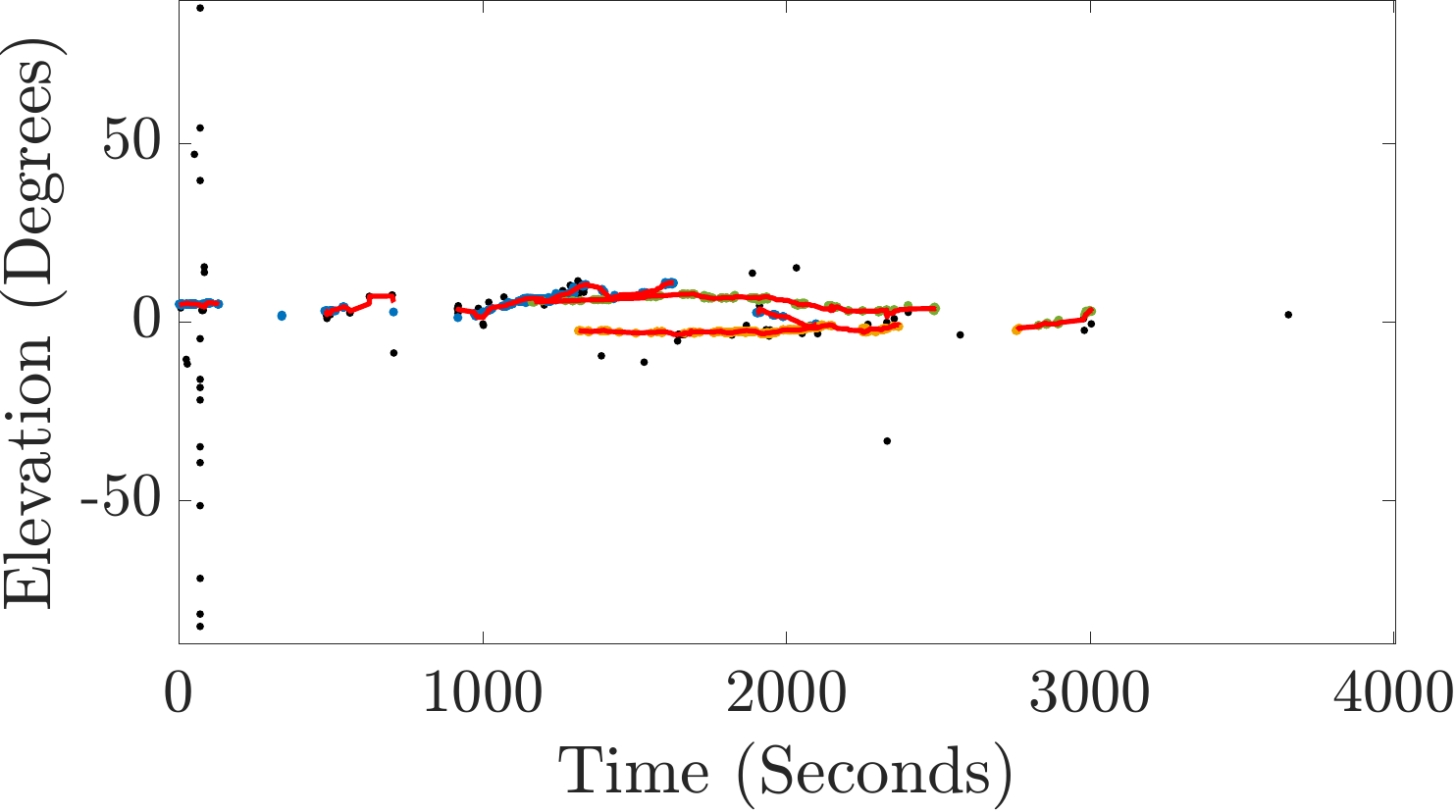}
        \subcaption{}
    \end{subfigure}
    
    \caption{{Tracking results obtained by \ac{bp}-based \ac{mtt} with track smoothing and stitching in scenarios with {one to three} whales, shown in azimuth and elevation versus time.}} 
    \label{fig:simpleencounter}
    \vspace{-3mm}
\end{figure*}

\begin{figure*}[t]
    \centering
    \begin{subfigure}{0.5\textwidth}
        \includegraphics[width=\linewidth, clip]{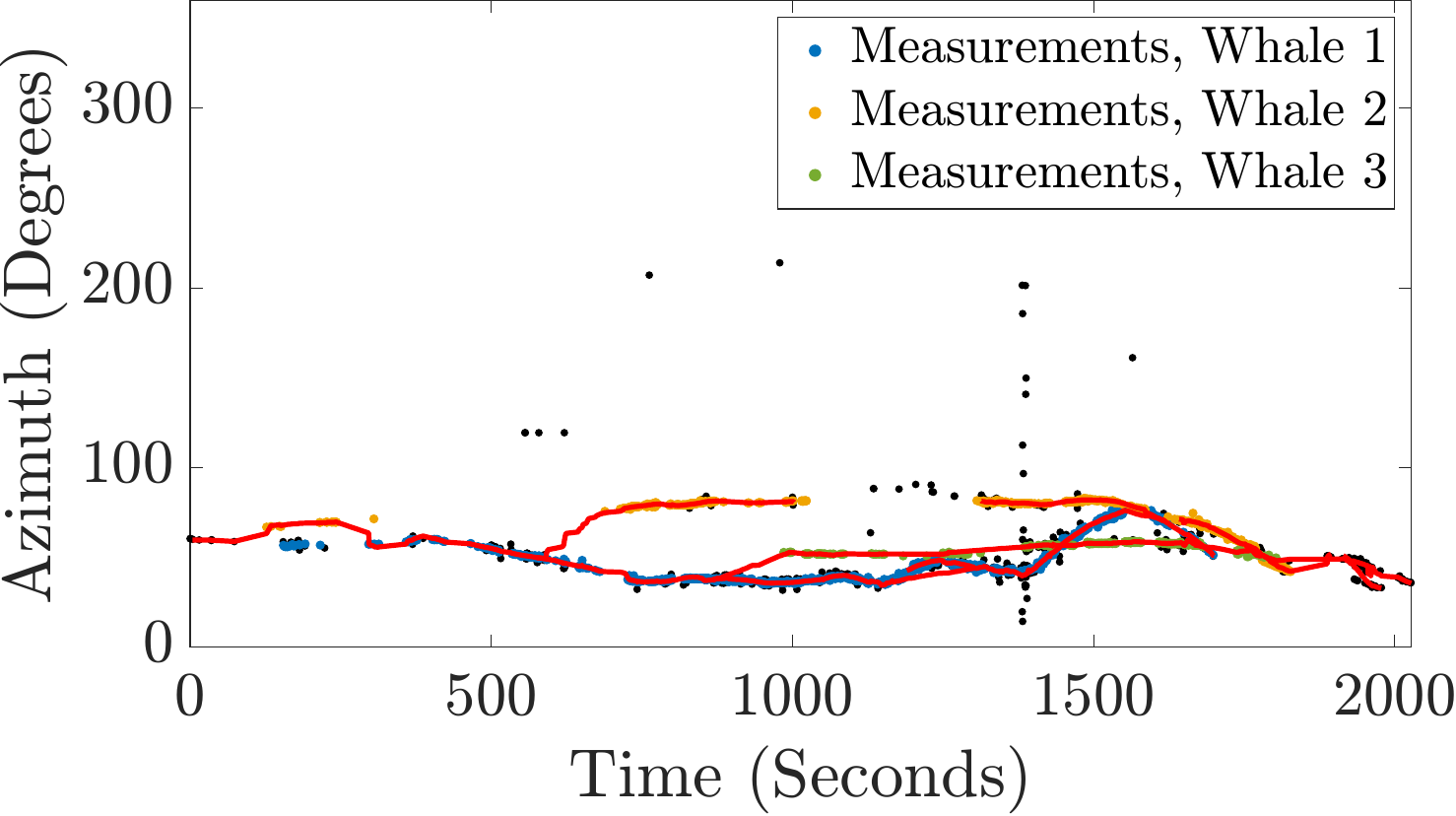}
        \subcaption{}
    \end{subfigure}%
        ~ 
    \begin{subfigure}{0.5\textwidth}
        \includegraphics[width=\linewidth, clip]{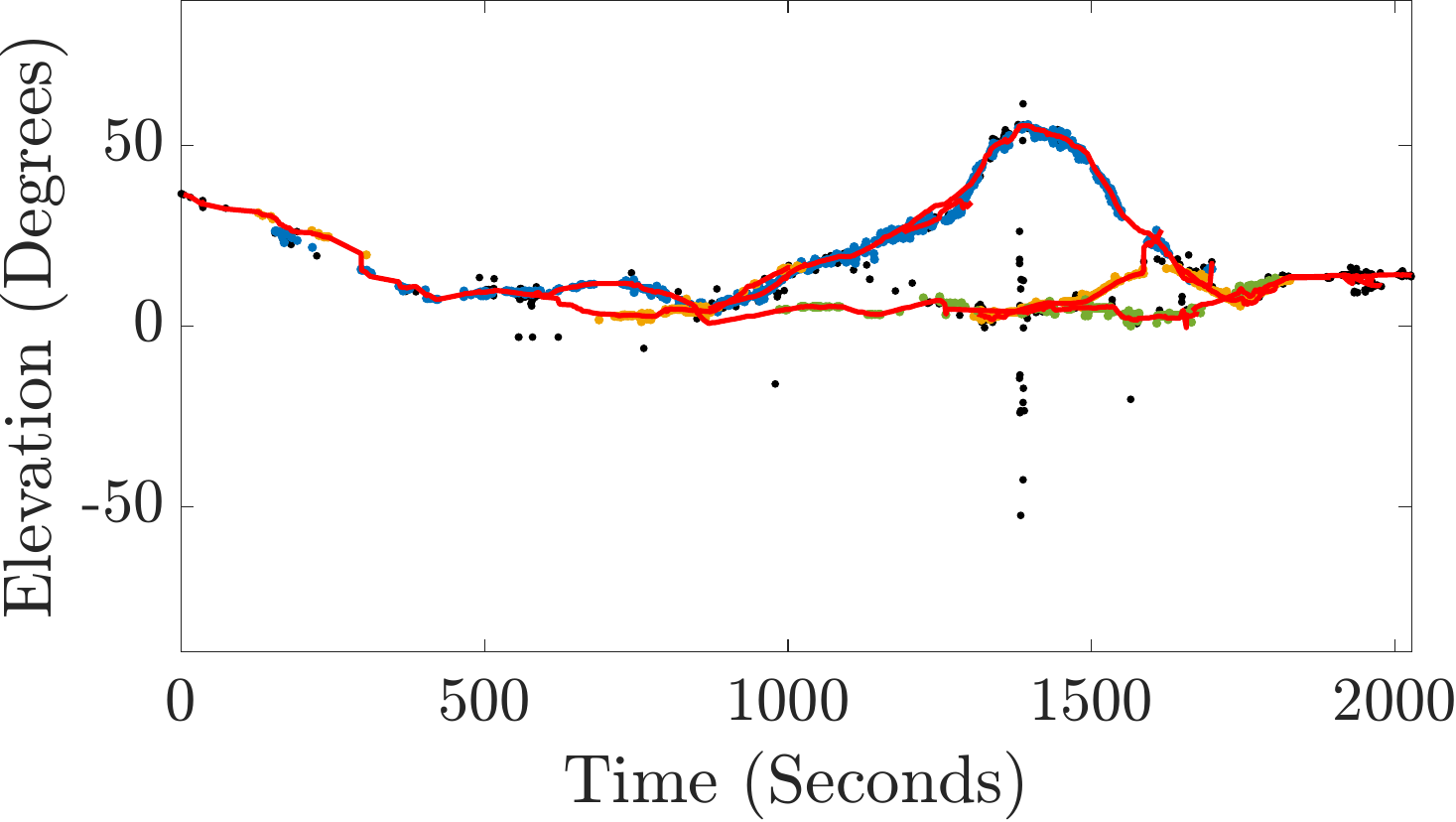}
        \subcaption{}
    \end{subfigure}

    \begin{subfigure}{0.5\textwidth}
        \includegraphics[width=\linewidth, clip]{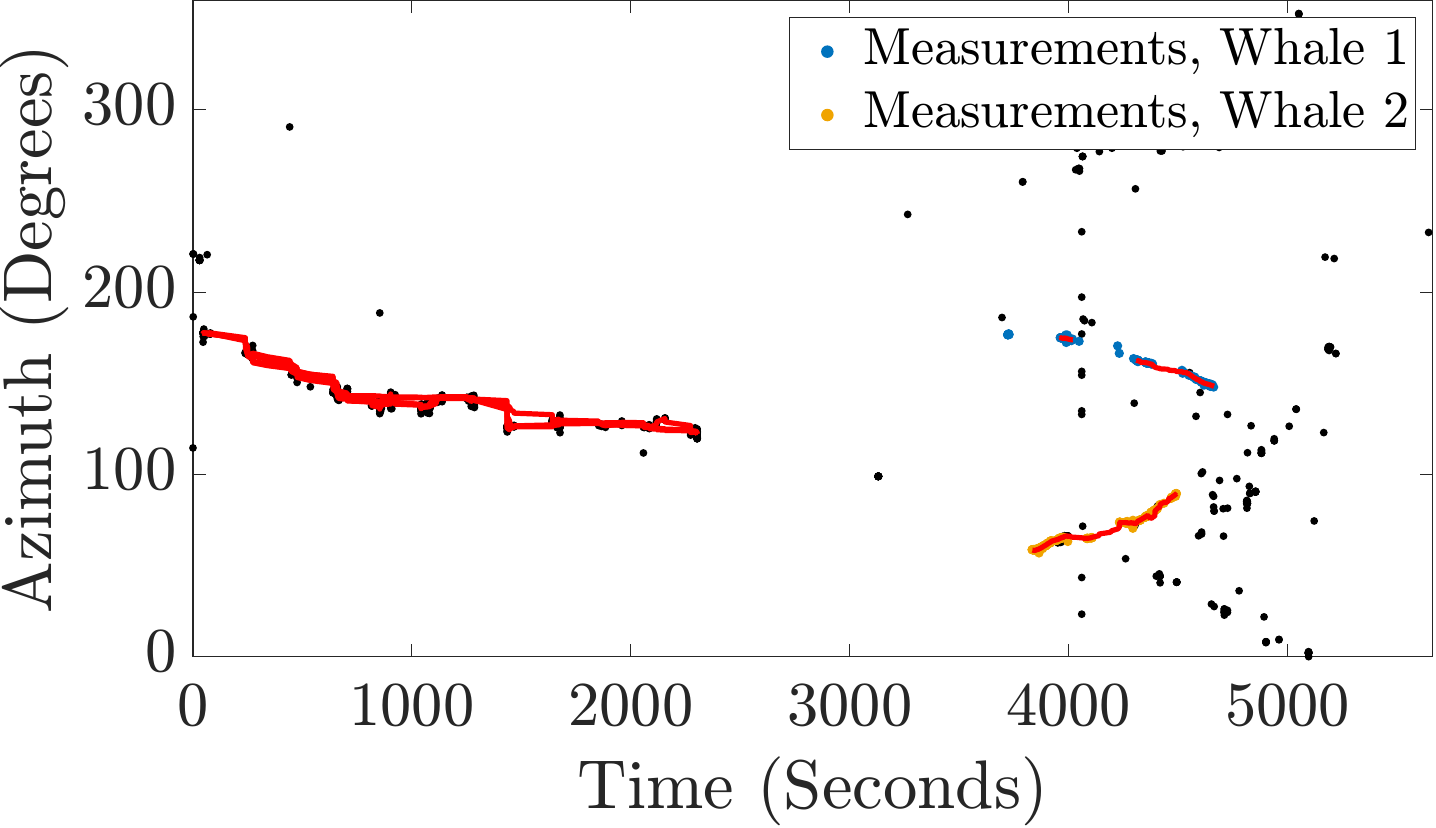}
        \subcaption{}
    \end{subfigure}%
        ~ 
    \begin{subfigure}{0.5\textwidth}
        \includegraphics[width=\linewidth, clip]{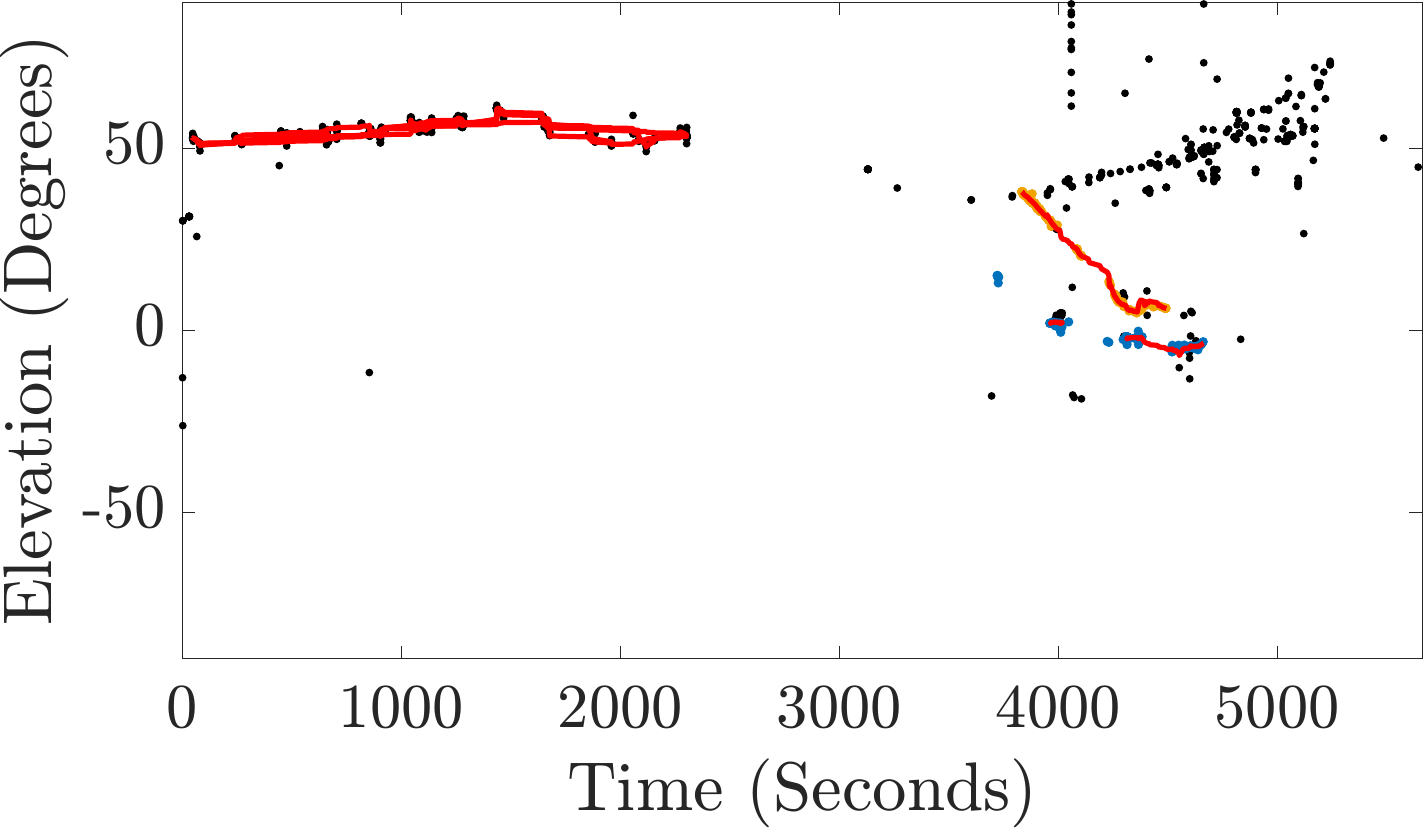}
        \subcaption{}
    \end{subfigure}
    
    \caption{{Tracking results obtained by \ac{bp}-based \ac{mtt} with track smoothing and stitching in more complex scenarios with multiple whales, shown in azimuth and elevation versus time.}} 
    \label{fig:complexencounter}
    \vspace{-3mm}
\end{figure*}

\begin{figure*}[t]
    \centering
    \begin{subfigure}{1\textwidth}
        \includegraphics[width=\linewidth, clip]{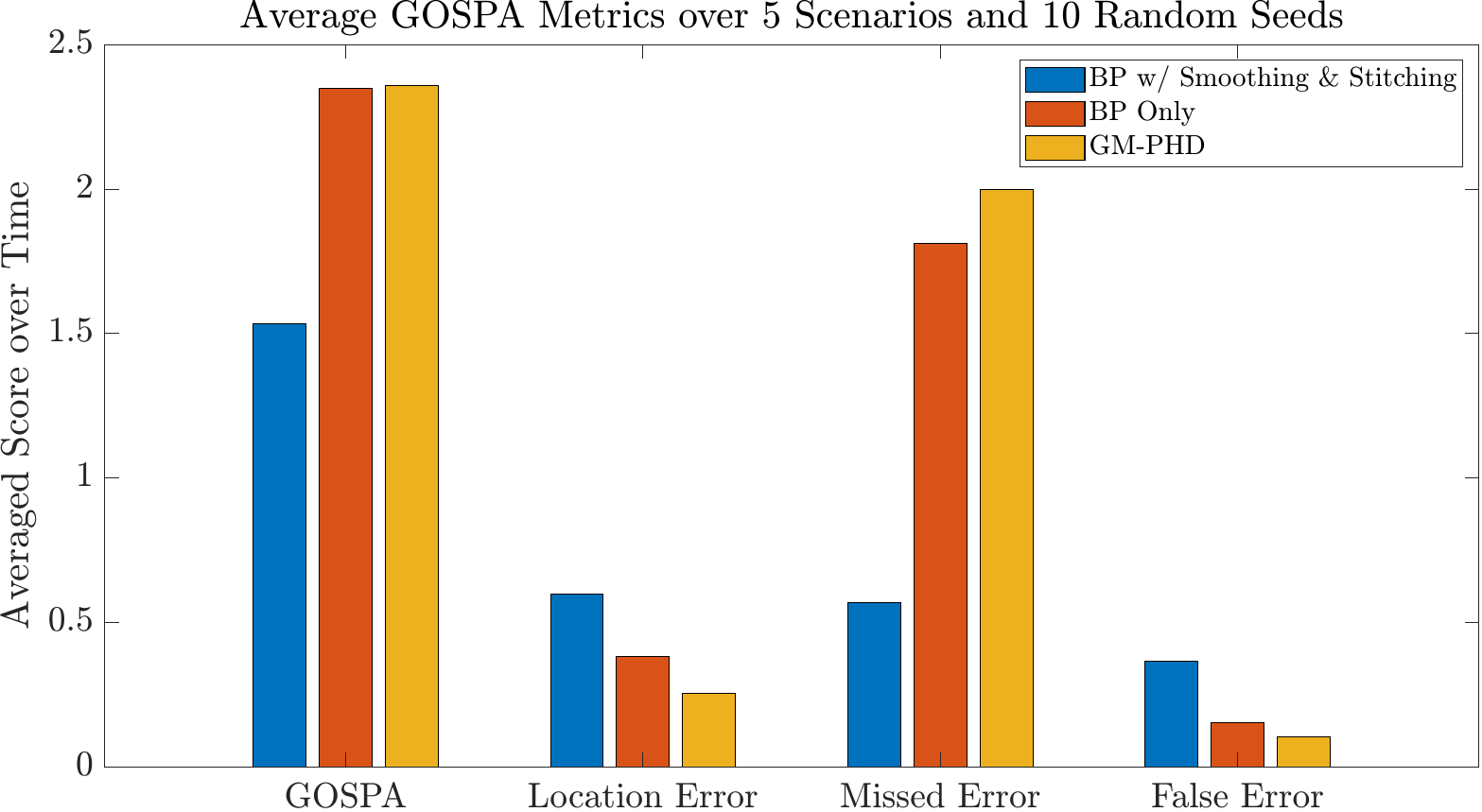}
        \subcaption{}
    \end{subfigure}%
    
    \caption{{Time-averaged \ac{gospa} metrics, subsequently averaged over the five scenarios shown in Figs.~\ref{fig:simpleencounter} and~\ref{fig:complexencounter} and over runs with random seeds 1--10.}} 
    \label{fig:gospareal}
    \vspace{-3mm}
\end{figure*}

In the following, each featured encounter includes localized \textit{Ziphius cavirostris} detections from one or more whales during encounters ranging $0.5$--$2$ hours which have been manually labeled\cite{BagSnyBerCurWigHilSirFraBau:J25}.

{The three encounters in Fig.~\ref{fig:simpleencounter} are relatively simple cases involving one to three whales, whose tracks are readily identifiable even by an inexperienced human operator. In each encounter, the tracks do not overlap in at least one of the two domains (azimuth or elevation), and clutter is sparse. Click detections occur in short bursts or trains, usually separated by small gaps, although pauses lasting several minutes occasionally occur. Our \ac{bp}-based \ac{mtt} method with smoothing and stitching produced tracks that closely followed the human-labeled clicks and resolved many of the small gaps between click sequences. However, larger gaps, particularly those lasting approximately $2$--$5$ min, remained unresolved.}

{The two encounters in Fig.~\ref{fig:complexencounter} illustrate the performance of our method in more complex cases. The upper encounter includes tracks from three whales that overlap in both azimuth and elevation, with more clutter near the target-generated detections. The \ac{bp}-based \ac{mtt} tracks again largely followed the human-labeled clicks, but the higher density of both target-generated and clutter detections caused two tracks to be created for the same whale. Our method was also more likely to make incorrect associations, during either initial tracking or stitching, when detection gaps coincided with crossings of multiple whale trajectories. Consequently, more gaps remained near track intersections, and some estimated tracks followed the wrong target after a gap. The lower encounter illustrates a case in which a trained human operator may still be required to verify the \ac{mtt} results. The clicks from two \textit{Ziphius cavirostris}, represented by blue and orange detections, were mostly tracked accurately by our method, although a few gaps remained unstitched. However, extraneous tracks were also generated for a non-beaked whale target. The click sequences observed from $0$ to $2400$ s most likely belong to a dolphin species rather than the target species. Such nonrandom clutter can readily produce false-positive tracks in \ac{mtt}.}

Fig.~\ref{fig:gospareal} summarizes the error metrics for these five encounters and compares \ac{bp}-based \ac{mtt} with smoothing and stitching against standalone \ac{bp}-based \ac{mtt} and the \ac{gmphd}. The mean total \ac{gospa} errors were $1.534$, $2.349$, and $2.358$, respectively. Smoothing and stitching reduced the total error of standalone \ac{bp}-based \ac{mtt} by approximately $34.7\%$, primarily by reducing the missed-target error from $1.811$ to $0.570$. This improvement was accompanied by increases in localization error from $0.383$ to $0.598$ and false-target error from $0.154$ to $0.367$, which can result from incorrect associations between track fragments during stitching. Standalone \ac{bp}-based \ac{mtt} also yielded a slightly lower mean total \ac{gospa} than the \ac{gmphd} ($2.349$ versus $2.358$) and had lower total error for nine of the ten seeds.

\section{Conclusion}\label{sec:conclusion}
{In this paper, we presented an automated method for tracking beaked whales in the \ac{doa} domain using passive acoustic recordings of echolocation clicks.} 
{Our approach extends \ac{bp}-based \ac{mtt} by introducing track smoothing and track stitching as post-processing steps that improve state estimation for complex data. The proposed method can (i) produce tracking results with lower error than other state-of-the-art methods and results comparable to reference labels provided by a trained human operator and (ii) bridge gaps caused by sequences of missed detections that violate standard \ac{mtt} assumptions. This approach may provide marine-mammal researchers with a useful tool that substantially reduces the effort required to process \ac{pam} data, enabling more effective studies of important species such as beaked whales.}
{Future work will demonstrate a fully automated tracking pipeline that extends our 2-D method to 3-D.}
\section{Acknowledgment}\label{sec:ack}
This work was supported in part by the National Science Foundation (NSF) under CAREER Award No. 2146261, the Office of Naval Research under Grants N00014-23-1-2284, N00014-20-1-4000, and  N00014-23-1-2435, as well as the Austrian Science Fund (FWF) under Grant J\,4726-N\vspace{.5mm}. 

\section{Author Declarations}
\subsection{Conflict of Interest}
The authors have no conflicts to disclose.

\renewcommand{\baselinestretch}{.99}
\selectfont
\bibliographystyle{IEEEtran}
\bibliography{IEEEabrv,StringDefinitions,SALPapers,SALBooks,Temp}

\end{document}